\documentclass[fleqn,usenatbib]{mnras}
\usepackage{newtxtext,newtxmath}
\usepackage[T1]{fontenc}
\usepackage{ae,aecompl}
\newcommand{\hMpc}{\ensuremath{~h^{-1}\mathrm{Mpc}}}
\newcommand{\hMsun}{\ensuremath{~h^{-1}\mathrm{M}_\odot}}

\usepackage{graphicx}	% Including figure files
\usepackage{amsmath}	% Advanced maths commands
\usepackage{amsmath}
\usepackage{hyperref}
\usepackage{booktabs}   % nice tables

\usepackage{bm}
\usepackage{mathtools}
\usepackage{esvect}
\usepackage{multicol}
\usepackage[export]{adjustbox}

\title[Ringing the universe with cosmic emptiness]{Ringing
the universe with cosmic emptiness: void properties through
a combined analysis of stacked weak gravitational and Doppler lensing
}

\author[Hossen et al.]{
Md Rasel Hossen$^{1}$\thanks{E-mail: mhos5289@uni.sydney.edu.au},
Sonia Akter Ema$^{1}$, 
Krzysztof Bolejko$^{2}$ and 
Geraint F. Lewis$^{1}$
\\
$^{1}$Sydney Institute for Astronomy, School of Physics, A28, The University of Sydney, NSW 2006, Australia \\
$^{2}$School of Natural Sciences, College of Sciences and Engineering, University of Tasmania, Private Bag 37, Hobart TAS 7001, Australia
}

\date{Accepted XXX. Received YYY; in original form ZZZ}

\pubyear{2022}

\begin{document}
\label{firstpage}
\pagerange{\pageref{firstpage}--\pageref{lastpage}}
\maketitle

% Abstract of the paper
\begin{abstract}
An essential aspect of cosmic voids is that these underdense regions provide complementary information about the properties of our Universe. Unlike dense regions, voids are avoided by matter and are less contaminated by baryonic processes. The first step to understanding the properties of cosmic voids is to correctly infer their mass profiles. In the literature, various techniques have been implemented. In this paper, we review them and implement a new technique that is based on Doppler lensing. We use a relativistic $N$-body code, \textsc{Gevolution}, to generate cosmological mass perturbations and implement a three-dimensional ray-tracing technique, which follows the evolution of a ray-bundles. We focus on the various properties of cosmic voids (e.g. void size function, 2-point correlation function, and the density profile of voids), and compare the results with their universal trends.
We show that when weak-lensing is combined with the Doppler lensing we obtain even tighter constraints than weak-lensing alone. We also obtain better agreement between density profiles within central parts of voids inferred from lensing and density profiles inferred from halo tracers. The implication of the result relevant to the ongoing and prospective low-redshift spectroscopic surveys is briefly discussed.
\end{abstract}

% Select between one and six entries from the list of approved keywords.
% Don't make up new ones.
\begin{keywords}
gravitational lensing: weak-- large-scale structure of Universe-- dark matter-- methods: numerical
\end{keywords}

%%%%%%%%%%%%%%%%%%%%%%%%%%%%%%%%%%%%%%%%%%%%%%%%%%

%%%%%%%%%%%%%%%%% BODY OF PAPER %%%%%%%%%%%%%%%%%%

\section{Introduction}

% Define void  and its history.
Cosmic voids -- the underdense regions -- form a significant portion of the cosmic web filling more than half of the volume of the observable Universe \citep{1994AJ....108..745V}. Because of their peculiar nature \citep{1985ApJS...58....1B, 1987ApJ...313..505W, 2001ApJ...557..495P}, significant attention has been paid to understanding the physics of cosmic voids since their first discovery in the late 1970s \citep{1978ApJ...222..784G, 1978MNRAS.185..357J}. The physics of cosmic voids introduces a new cosmic probe in understanding the evolution of our Universe \citep{2019BAAS...51c..40P, 2022arXiv220107241M}. Recent advancements in both galaxy redshift surveys \citep{2013AJ....145...10D, 2005astro.ph.10346T, 2011AJ....142...72E, 2013ExA....35...25D} and cosmological simulations \citep{2005MNRAS.364.1105S, 2016MNRAS.463.1797D, 2017ComAC...4....2P, 2020ApJS..250....2V} have attracted scientists to focus on voids and their relevance in the evolution of large-scale structures (LSS) to gain insights into our Universe \citep{2006MNRAS.372.1710P, 2007MNRAS.375..489H, 2014PhRvD..90j3521C, 2020MNRAS.493..899H, 2021ApJ...919...24B}.

The void abundance, also known as the void size function, is one of the features of cosmic voids, that is particularly sensitive to the dark energy equation of state \citep{2009ApJ...696L..10L, 2012MNRAS.426..440B, 2015PhRvD..92h3531P} and offers a new framework for exploring dark energy and modified gravity \citep{2009ApJ...696L..10L, 2012MNRAS.426..440B, 2013PhRvL.111x1103S,  2015MNRAS.451.1036C, 2015PhRvD..92h3531P, 2015MNRAS.451.4215Z, 2016PhRvL.117i1302H, 2018MNRAS.475.3262F, 2019A&A...632A..52P, 2019JCAP...12..040V}. 
\citet{2016ApJ...820L...7S} demonstrated that the combined analysis of the abundance of voids and clusters might break the degeneracy among cosmological parameters and provide tighter constraints on modified gravity using extreme value statistics. 
In recent decades, researchers have also concentrated their efforts on extracting cosmological information by examining various aspects of cosmic voids, including the void-galaxy auto/cross-correlation function \citep{2013MNRAS.436.3480P, 2014PhRvL.112d1304H, 2015JCAP...11..036H, 2016MNRAS.459.4020L, 2016MNRAS.462.2465C, 2017MNRAS.469..787P, 2017PhRvD..95h3502A, 2018PhRvD..98d3527N, 2019MNRAS.483.3472N, 2019PhRvD.100b3504N, 2020ApJ...901...87P, 2022MNRAS.511.4333K}, and the universal density profile \citep{2012ApJ...754..109L, 2012ApJ...761...44S, 2012ApJ...761..187S, 2013ApJ...762L..20K, 2014MNRAS.443.3238P, 2014MNRAS.440..601R, 2014PhRvL.112y1302H, 2015JCAP...03..047L, 2015MNRAS.449.3997N, 2015JCAP...08..028B, 2017MNRAS.469..787P, 2020ApJ...901...87P, 2021arXiv210913378R, 2021PhRvD.104b3512W}, which could serve as a potential probe for studying cosmology and galaxy formation with greater precision.

Weak gravitational lensing is defined as the distortion of the size (in the form of convergence, $\kappa$) and shape (in the form of shear, $\gamma$) of background galaxies by the mass distribution 
along the line of sight. Thus, it can be directly used to map the distribution of dark matter \citep{1992ApJ...388..272K, 2000Natur.405..143W, 2001PhR...340..291B, 2008ARNPS..58...99H}. Weak-lensing by cosmic voids, also known as void lensing, has received a lot of attention in recent decades \citep{1999MNRAS.309..465A, 2013ApJ...762L..20K, 2013MNRAS.432.1021H, 2015MNRAS.451.1036C, 2015JCAP...08..028B, 2018PhRvD..98b3511B, 2018MNRAS.480L.101D, 2021MNRAS.507.2267D}. The weak-lensing tangential shear signal around cosmic voids has recently been investigated \citep{2014MNRAS.440.2922M, 2015MNRAS.454.3357C, 2017MNRAS.465..746S} and reported to be moderately significant ($4.4-7 \sigma$) while \citet{2019MNRAS.490.3573F} found the most significant detection ($14 \sigma$). Subsequently, \citet{2016MNRAS.455.3367G} and \citet{2018MNRAS.481.5189B} utilised a new approach to quantify the tangential shear signal with higher significance ($10-15 \sigma$). Besides the weak-lensing signal by cosmic voids, the lensing imprint of cosmic voids on cosmic microwave background \citep{2016PhRvD..93d3523C,2020ApJ...890..168R, 2021MNRAS.500..464V, 2022arXiv220311306K} and intrinsic alignments of galaxies around cosmic voids \citep{2022MNRAS.509.1985D} have recently been studied.

Doppler lensing is an alternative approach to gravitational lensing that originates from redshift-space distortion \citep[arises due to the peculiar velocities of galaxies;][]{1987MNRAS.227....1K, 2013MNRAS.436.3480P, 2016MNRAS.462.2465C, 2017JCAP...07..014H, 2020JCAP...12..023H, 2017A&A...607A..54H, 2019PhRvD.100l3513A, 2022MNRAS.509.1871C} and is represented in terms of weak-lensing \citep{2013PhRvL.110b1302B, 2014MNRAS.443.1900B}. As redshift distortions are caused by the underlying matter distribution so, it could be beneficial to examine Doppler lensing in conjunction with weak-lensing to improve the precision of lensing measurements. \citet{2014MNRAS.443.1900B} measured the Doppler lensing signal in combination with the weak-lensing signal by using a Newtonian simulation. They compute the 2-point auto/cross-correlation function and observed tighter constraints for cosmological parameters. They also showed that the Doppler lensing dominates over gravitational lensing at medium-to-low redshifts, leading to a new probe to consider in conjunction with weak-lensing for greater precision to study the underlying matter distribution in and around cosmic structures. Subsequently, \citet{2017MNRAS.472.3936B} investigated the effect of Doppler magnification on the size of galaxies and constructed an estimator to measure the velocities using this effect. In our recent paper \citep{2022MNRAS.509.5142H}, we studied both weak-lensing and Doppler lensing signals in and around the cosmic voids/haloes and demonstrated that the most optimal technique for mapping the mass distribution that combines both gravitational and Doppler lensing effects should target the redshift range $z\approx 0.3-0.4$.

In this paper, we investigate further benefits of combining the gravitational and Doppler lensing methods. This is done by the means of numerical simulations.
Numerical simulations are essential for developing a deeper understanding of the LSS in the universe as it becomes complex on the nonlinear scale, and scientists have published a large number of works on detecting weak-lensing signals by the mass distribution of LSS using Newtonian simulations \citep{2000ApJ...530..547J, 2011ApJ...742...15T, 2012A&A...541A.161V}. But in the presence of relativistic sources, Newtonian simulation has some limitations, and one may need to consider relativistic simulations to capture all the relativistic effects \citep{2012PhRvD..85f3512G}. In this work, we will consider a relativistic approach to cosmological simulations developed by \citet{2016NatPh..12..346A, 2016JCAP...07..053A}.

In this paper, we measure the lensing signals around cosmic voids and explore various properties of cosmic voids (e.g. void size function, 2-point auto/cross-correlation function, and density profile of voids). We constrain the lensing model parameters from the theoretical predictions and numerical data to see how much constraining power we can enhance by considering solely weak-lensing shear or magnification signal and combined lensing (weak-lensing + Doppler lensing) signals. In the following, we outline the theoretical background and numerical recipes in Section \ref{background}, describe the results of the various properties of cosmic voids in Section \ref{void_porperties}. The implications of our findings of the lensing measurements are then discussed in Section \ref{lensing_measurements} and the drawn conclusions are summarised in Section \ref{conclusions}.

\section{Background and Methodology}\label{background}
\subsection{Lensing theory}
\subsubsection{Weak gravitational lensing}
Gravitational lensing is a unique technique for directly mapping the mass distribution of dark matter throughout the cosmos. Weak gravitational lensing systematically distorts the images of source galaxies by mapping them to new locations on the sky. The lens equation for a gravitationally lensed image can be expressed as
\begin{equation}
    \pmb{\alpha} = \pmb{\beta} - \pmb{\theta} \, ,
\end{equation}
where $\pmb{\alpha}$ is the deflection angle,   $\pmb{\beta}$ is the angular position of the source on the sky, and $\pmb{\theta}$ is the observed position of the lensed image. Following \citet{2001PhR...340..291B}, the deformation matrix $\mathcal{A}$ is the equivalent Jacobian matrix of the lens mapping and can be written as 
\begin{equation}
    \mathcal{A}_{ij} = \frac{\partial \beta_{i}}{\partial \theta_{j}} = \delta_{ij} - \frac{\partial \alpha_{i}}{\partial \theta_{j}} \, .
    \label{eq:matrix}
\end{equation}
This deformation matrix can be parameterised using more physically instructive notions such as convergence ($\kappa$) and shear ($\gamma\equiv \gamma_1+i\gamma_2$), and it can be represented as
\begin{align}
%\mathcal A(\vec{\theta})
\mathcal A=\left( \begin{array}{c c}
1-\kappa-\gamma_1 & -\gamma_2 \\
-\gamma_2 & 1-\kappa+\gamma_1\\
\end{array} \right).
\end{align}
And, the total lensing magnification $\mu$ is expressed by the determinant of the inverse matrix
\begin{equation}
\mu = \frac{1}{\det(\mathcal A)}=[(1-\kappa)^2-|\gamma|^2]^{-1}.\label{mag}
\end{equation}
The convergence describes the magnification and de-magnification of source images, whereas the shear describes how much the image is stretched along its axis. Depending on the mass distribution of the universe, the convergence along the line of sight can be positive, implying a magnified image of the source, or negative, implying a de-magnified image of the source. The positive or negative value of the convergence indicates that the photon travels through an overdense or underdense area of the universe. The weak-lensing statistical quantities i.e. convergence, shear, and magnification are also related to the projected surface mass density ($\Sigma$) and differential surface mass density of the lens ($\Delta\Sigma$) and can be expressed as
\begin{equation}\label{eq:proj} 
    \kappa = \frac{\Sigma(r)}{\Sigma_{\rm cr}}~, \quad  \mu \approx 1 + 2\frac{\Sigma(r)}{\Sigma_{\rm cr}}, \quad {\rm and} \quad \gamma =\: \frac{\Delta\Sigma(r)}{\Sigma_{\rm cr}},
\end{equation}
where  $r$ is the physical transverse distance on the lens plane
\begin{equation}
    \Delta\Sigma(r) =\: \bar\Sigma(< r) - \Sigma(r),
\end{equation}
and $\Sigma_{\rm cr}$ is written as 
\begin{equation}
    \Sigma_{\rm cr} = \frac{c^2}{4 \pi G} \frac{D_{\rm os}}{D_{\rm ol} D_{\rm ls}},
    \label{critical}
\end{equation}
where $D_{ij}$ are the angular diameter distances between the observer $(\rm o)$, lens $(\rm l)$, and source $(\rm s)$.

\subsubsection{Doppler lensing} \label{dop}
Doppler lensing is the apparent change in object size and magnitude due to peculiar velocities. There are two fundamental differences between standard weak gravitational lensing and Doppler lensing. Firstly, weak-lensing is the phenomenon in which the mass distribution along the line of sight influences both the size and shape of the source galaxies, whereas Doppler lensing affects only the size of the galaxies while leaving their shape unaltered. Secondly, the largest contribution to the weak-lensing signal comes from the matter that is approximately located in mid-distance between the observer and the observed galaxies, and consequently, the precise location of the source galaxies is less significant. In contrast, Doppler lensing is the local consequence of the observed structure, thus the source is at the lens. 
The reason why for the Doppler lensing the precise location of the source is important is that the effect is caused by the peculiar motion of galaxies \citep{2008PhRvD..78l3530B,2013PhRvL.110b1302B, 2014MNRAS.443.1900B, 2017MNRAS.472.3936B}. The convergence sourced by the Doppler lensing is given by following normalised equation\footnote{we use the Hubble parameter with a unit of $h^{-1}\rm Mpc$, so the product of $H$ and $\chi_s$ is a dimensionless quantity.}
\begin{equation}
    \kappa_v=\left(1-\frac{1+z_s}{ H\chi_s}\right) \left(\frac{\bm{v}\cdot\bm n}{c} \right),
    \label{eqn:Dop}
\end{equation}
where $\kappa_v$ is the Doppler convergence, $z_s$ is the source redshift, $\chi_s$ is the co-moving distance of the sources, $H$ is the Hubble parameter, \bm{$v$} is the velocity between source galaxies and the observer, $c$ is the speed of light, and \bm{$n$} is the unit vector from the source to the observer. There are two scenarios can be observed because of the propagation direction of the photon
\begin{itemize}
    \item ${\bm{v}\cdot\bm n} > 0$, implying that the objects are moving towards us, thus the Doppler convergence $\kappa_v < 0$. This condition suggests that the observed structures will be measured smaller and dimmer than the usual objects at their observed redshift.
    
    \item ${\bm{v}\cdot\bm n} < 0$, implying that the objects are moving away from us, thus the Doppler convergence $\kappa_v > 0$. This condition suggests that the observed structures will be measured larger and brighter than the usual objects at their observed redshift.
\end{itemize}
Due to the factor $(H \chi_s)^{-1}$ in Eqn. \ref{eqn:Dop},  the Doppler lensing convergence decreases with redshift, which limits its application to low-redshift only.

\subsection{Numerical recipes}
\subsubsection{\texorpdfstring{$N-$}-body simulation}

In this paper, we use relativistic $N$-body code \textsc{Gevolution}\footnote{{\textcolor{blue}{https://github.com/gevolution-code/gevolution-1.2}}} \citep{2016NatPh..12..346A, 2016JCAP...07..053A} to generate the weak perturbations, in which the perturbed Friedmann-Lemaître-Robertson-Walker line element of the metric has the form
\begin{align}\label{e:metric}
    ds^2 = a^2(\tau) [-(1+2\Psi)d\tau^2 - 2B_ix^id\tau + (1-2\Phi)\delta_{ij}dx^idx^j \notag\\
    + h_{ij}dx^idx^j],
\end{align}
where $a$ represents the scale factor of the background, $x^i$ are the comoving Cartesian coordinates, and $\tau$ is the conformal time. $\Phi$ and $\Psi$ are the scalar perturbations, $B_i$ is the vector perturbation and $h_{ij}$ is the tensor perturbation. To minimise computational time and memory, we consider only scalar ($\Phi$ and $\Psi$) perturbations here since vector and tensor perturbations have the negligible effect on light propagation. For more details we refer the reader to these papers \citep{2020MNRAS.497.2078L, 2022MNRAS.509.3004E}.    

The \textsc{Gevolution} simulation consists in a baseline of standard $\Lambda$ cold dark matter ($\Lambda$CDM) model and we run it by using the following cosmological parameters: $h = 0.67556$, $\Omega_{\rm c} = 0.2638$, $\Omega_{\rm b} = 0.048275$, and a radiation density that includes massless neutrinos with $N_\mathrm{eff} = 3.046$. Our considered CDM simulation follows the evolution of $256^3$ CDM particles in a cubic comoving volume $(320~\hMpc)^3$ from redshift $z_\mathrm{ini} = 127$ to the present epoch, and the linear initial conditions are computed with \textit{CLASS} \citep{2011JCAP...07..034B} by assuming a primordial power spectrum with amplitude $A_s = 2.215 \times 10^{-9}$ and spectral index $n_s = 0.9619$. This provides a spatial resolution of $1.25~\hMpc$, and corresponds to a minimum resolved halo mass (with 20 particles) of $5 \times 10^{11}~\hMsun$. We run \textsc{Gevolution} simulation for 21 realisations (by changing the initial conditions but keeping the identical cosmological parameters), which takes around 250 CPU hours to provide the output in the forms of particle snapshots and weak potentials.

\subsubsection{Void finder}
Cosmic voids are the under-dense patches of the universe surrounded by filaments and sheets of galaxies that make up the cosmic web. There are a variety of void-finding techniques in the literature, including Voronoi tessellation and watershed methods \citep{2007MNRAS.380..551P, 2008MNRAS.386.2101N, 2012ApJ...761...44S, 2019PhRvD.100b3504N}, Delaunay density estimation methods \citep{2000A&A...363L..29S, 2016MNRAS.459.2670Z}, 2D projections \citep{2015MNRAS.454.3357C}, Hessian-based methods \citep{2017MNRAS.468.3381A, 2018MNRAS.479.4861A}, tunnel methods \citep{2018MNRAS.476.3195C, 2021MNRAS.507.2267D}, spherical underdensity estimation methods \citep{2005MNRAS.360..216C, 2005MNRAS.363..977P, 2011MNRAS.411.2615L, 2012MNRAS.421..926P, 2012MNRAS.426.3041H, 2018ApJ...866..135V}. We recommend the readers to these papers \citep{2008MNRAS.387..933C, 2018MNRAS.476.3195C} for further information on the comparison of void finding methodologies. Here we employ \textsc{Revolver} (REal-space VOid Locations from surVEy Reconstruction)\footnote{{\textcolor{blue}{https://github.com/seshnadathur/Revolver}}}, a publicly available code, to find the cosmic voids. It is a three-dimensional (3D) void finder that is been utilised extensively in simulations as well as observed catalogues \citep{2019PhRvD.100b3504N, 2019MNRAS.482.2459N, 2020ApJ...890..168R, 2020arXiv200201689L, 2021MNRAS.505.4626J, 2022MNRAS.511.4333K}. The methodology of the \textsc{Revolver} is based on the parameter-free void-finding algorithm \textsc{Zobov} \citep[ZOnes Bordering On Voidness;][]{2008MNRAS.386.2101N, 2015A&C.....9....1S}, which uses Voronoi tessellation to estimate local density minima from the input tracers.

The code \textsc{Revolver} assumes the centre of the void is the centre of the largest empty sphere that can be inscribed in the void, and consequently the place of the lowest density of the galaxy distribution \citep{2015MNRAS.454.2228N}. The barycentre from watershed voids has been extensively used in literature because it is more stable against shot noise and is dependent on the entire geometry of the cosmic void \citep{2015JCAP...11..036H, 2016PhRvL.117i1302H, 2019A&C....27...53C, 2020JCAP...12..023H}. However, the main results (lensing results) of this work, which will be discussed in Section \ref{lensing_measurements}, are not affected by the various assumptions about the centre of the cosmic voids. The ray-tracing algorithm is not sensitive to such a definition (it is affected by the matter distribution along the line of sight, not by any artificial definition of the centre of the void). Though realistic void shapes are far from spherical \citep[for a nice visualisation we refer the reader to Fig. 3 of][]{2016MNRAS.461..358N} but for simplicity the \textsc{Zobov} algorithm computes the effective void radius $R_{\rm void}$ from the total volume of each void (which is effectively the sum of the volumes of its constituent Voronoi cells, $V_{\rm void} = \sum_iV_i$) as
\begin{equation}\label{eq:Rv}
    R_{\rm void} = \left(\frac{3}{4\pi}V_{\rm void}\right)^{1/3}.
\end{equation}
The \textsc{Zobov} algorithm also computes the minimum galaxy density contrast near cosmic voids using the equation $\delta_{g, \rm min}= n_{g, \rm min}/\Bar{n} - 1$, where $n_{g, \rm min}$ is the minimum galaxy number density and $\Bar{n}$ is the average matter density. It is also called as core density contrast of each void from the estimation of Voronoi tessellation field estimator (VTFE) reconstruction.

The \textsc{Revolver} algorithm takes a discrete tracer distribution as an input, in this instance DM haloes or particles. Haloes are identified by using the publicly available code \textsc{Rockstar} \citep{2013ApJ...762..109B}, a phase-space friends-of-friends (FOF) halo finder. We feed the DM particle snapshots from 21 distinct \textsc{Gevolution} simulations at $z=0.22$ into \textsc{Rockstar}\footnote{{\textcolor{blue}{https://bitbucket.org/gfcstanford/rockstar/src/main/}}}, which provides on average $3 \times 10^4$ DM haloes per realisation. The redshift of $z = 0.22$ was chosen on the basis of the relevance of the Doppler lensing, which dominates for $z<0.3$ (see \citet{2014MNRAS.443.1900B}). A detailed explanation also given in the conclusion part of our recent paper (\citet{2022MNRAS.509.5142H}). To find DM haloes, we set the FOF refinement fraction = 0.7 and 3D-FOF linking length = 0.28, for more details we refer the reader to the original paper \citep{2013ApJ...762..109B}. Then we utilise DM haloes as an input tracer into \textsc{Revolver} to identify voids, which provide on average 210 cosmic voids per realisation with a radius spanning from $10~\hMpc-65~\hMpc$.

%%%%%%%%%%% FIGURE %%%%%%%%%%%%%
\begin{figure*}
    \centering
    %\hspace*{-0.6in}
    \includegraphics[width=2\columnwidth]{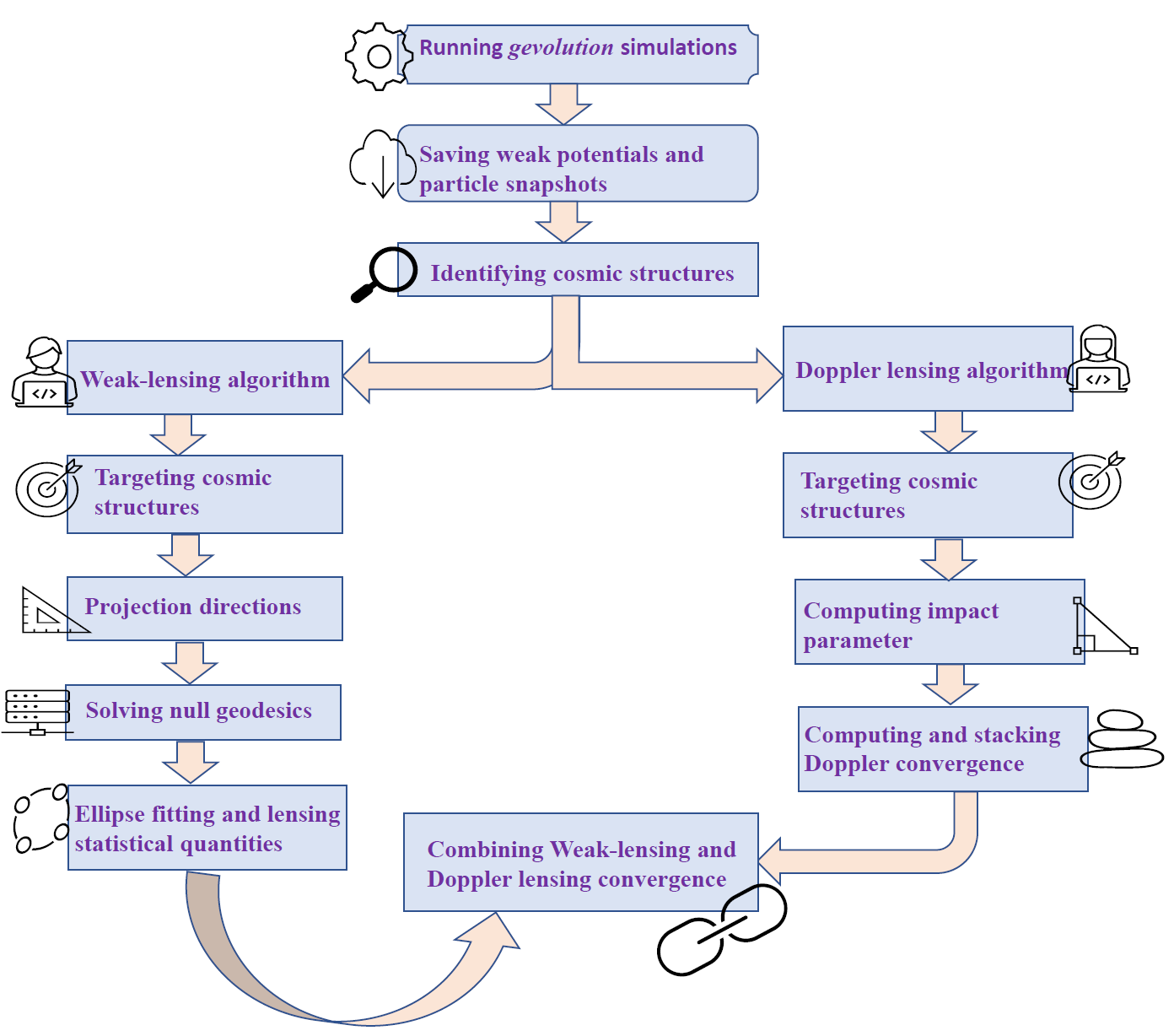}
    %\vspace*{-12mm}
    \caption{A flowchart diagram summarising the algorithm of combining the gravitational lensing and Doppler lensing to map the matter distribution within cosmic voids.}
    \label{flowchart}
\end{figure*}
%%%%%%%%%%%%%%%%%%%%%%%%%%%%%%%%

Running the VTFE technique on a large number of DM particles is computationally challenging, so we utilise the down-sampling routine to randomly sub-sample the DM particles from the output of 21 distinct \textsc{Gevolution} simulations to minimise the computational time and memory requirements, which is common in the literature \citep{2014MNRAS.442..462S, 2015MNRAS.454.2228N, 2022MNRAS.511.4333K}. In the down-sampling routine for each realisation of \textsc{Gevolution} simulation, we keep the DM particles down to a constant average density of $1.68 \times 10^6$ particles per cubic box size $(320~\hMpc)^3$, which equates to 10\% of the DM particles. Then we use the sub-sample of the DM particles as an input tracer into \textsc{Revolver} to identify voids, which provide on average 9315 cosmic voids per realisation with a radius spanning from $2~\hMpc-22~\hMpc$. It is worth noting that the void sizes and counts vary depending on the average density of the various input tracers.

\subsection{Lensing algorithms}
\subsubsection{Ray-tracing algorithm: weak gravitational lensing}

The most extensively utilised tool for precisely modelling and extracting gravitational lensing characteristics is ray-tracing techniques. Various approaches have been used in recent decades to examine gravitational lensing properties and other cosmological aspects, including optical scalar methods \citep{1991ApJ...370..481W, 1997PASJ...49..151N, 1999MNRAS.302..801H, 2019PhRvD.100b1301A}, statistical methods \citep{1991MNRAS.251..600B, 1992ApJ...388..272K, 1997ApJ...484..560J, 1999MNRAS.305..746M}, multiple-lens-plane methods: usually ray shooting methods \citep[RSM;][]{1989ApJ...337..581P, 1990ApJ...357...32L, 1990ApJ...365...22J, 2000ApJ...530..547J, 2009A&A...499...31H}, null geodesic methods \citep{1998PThPh.100...79T, 2012MNRAS.420..155K, 2020MNRAS.497.2078L}, ray bundle methods \citep[RBM;][]{1999MNRAS.306..567F, 2002MNRAS.331..180F, 2011MNRAS.416.1616F, 2019MNRAS.483.2671B, 2021A&A...655A..54B}, tree-based methods \citep{2007MNRAS.376..113A}, analytical methods \citep{2011MNRAS.415..881L}.

In this paper, we use \texttt{3D Ray Bundle Tracers} \citep[3D-RBT;][]{2022MNRAS.509.3004E, 2022MNRAS.509.5142H}, the 3D ray-tracing algorithm that relies on the design of the RBM \citep{1999MNRAS.306..567F, 2002MNRAS.331..180F}. The steps of the ray-tracing algorithm are briefly described below, however for more details on 3D-RBT, we refer the reader to our recent works \citep{2022MNRAS.509.3004E, 2022MNRAS.509.5142H}:
\begin{itemize}
    \item \textit{Generating weak perturbations:} the first step is to generate relativistic weak potentials using $N$-body simulation. We run 21 \textsc{Gevolution} simulations with the same cosmological parameters but different initial conditions and save the weak potentials and particle snapshots for different redshifts.
    
    \item \textit{Finding cosmic structures:} from the particle snapshots of \textsc{Gevolution} simulations at redshift $z=0.22$, we identify DM haloes using the publicly available code \textsc{Rockstar}. Then we utilise DM haloes as an input tracer into \textsc{Revolver} to identify cosmic voids.  
    
    \item \textit{Targeting cosmic structures:} after finding cosmic voids we subdivide them into three distinct radius ranges: i) $10~\hMpc \leq R_{\rm void} < 20~\hMpc$ ii) $20~\hMpc \leq R_{\rm void} < 30~\hMpc$, and iii) $30~\hMpc \leq R_{\rm void} < 40~\hMpc$.
    
    \item \textit{Projection directions:} we fix the observer at $z = 0$ and project bundles of photons (initially circular shape) in the directions of cosmic voids with three distinct groups of voids radii by slightly changing the projection angles.
    
    \item \textit{Solving null geodesics:} at each step of the integration we obtain the metric (\ref{e:metric}) from \textsc{Gevolution}. This is then used to evaluate the Christoffel symbols and obtain the direction of the path of the photon, and consequently the full bundle. The RBM implemented in this paper uses eight null geodesics and a central null geodesic per bundle. In addition to this at each step of integration we check the null condition. This is an essential sanity check to ensure that we are obtaining the correct photon path profile. 
    
    \item \textit{Ellipse fitting and lensing statistical quantities:} the initial circular shape of the bundle deforms as photons traverse through the cosmic web. The shape of the bundle is calculated by fitting an ellipse into the bundle's shapes. This is then used to compute the weak-lensing magnification and shear as
    \begin{eqnarray}\label{mu_eq}
        \mu = \frac{A_{i}}{A_{s}} \quad {\rm and} \quad \gamma = \frac{d-a}{d+a},
    \end{eqnarray}
    where $A_{i}$ represents the area of the image, $A_{s}$ is the area of the source, a \& d are the semi-major and semi-minor axes of an ellipse, respectively. The weak-lensing convergence ($\kappa$) is then  calculated from Eqn. \ref{mag}, and is
    \begin{eqnarray}\label{kappa_eq}
        \kappa = 1 - \sqrt{\frac{1}{\mu} + \gamma^2}.
    \end{eqnarray}
\end{itemize}

It is worth noting that ray-tracing algorithms are computationally challenging, requiring significant amounts of time and memory. For technical details and challenges of implementation of these algorithms we refer the reader to our earlier works \citep{2022MNRAS.509.3004E, 2022MNRAS.509.5142H}.

%%%%%%%%%%% FIGURE %%%%%%%%%%%%%
\begin{figure*}
    \centering
    %\hspace*{-0.6in}
    \includegraphics[width=2\columnwidth]{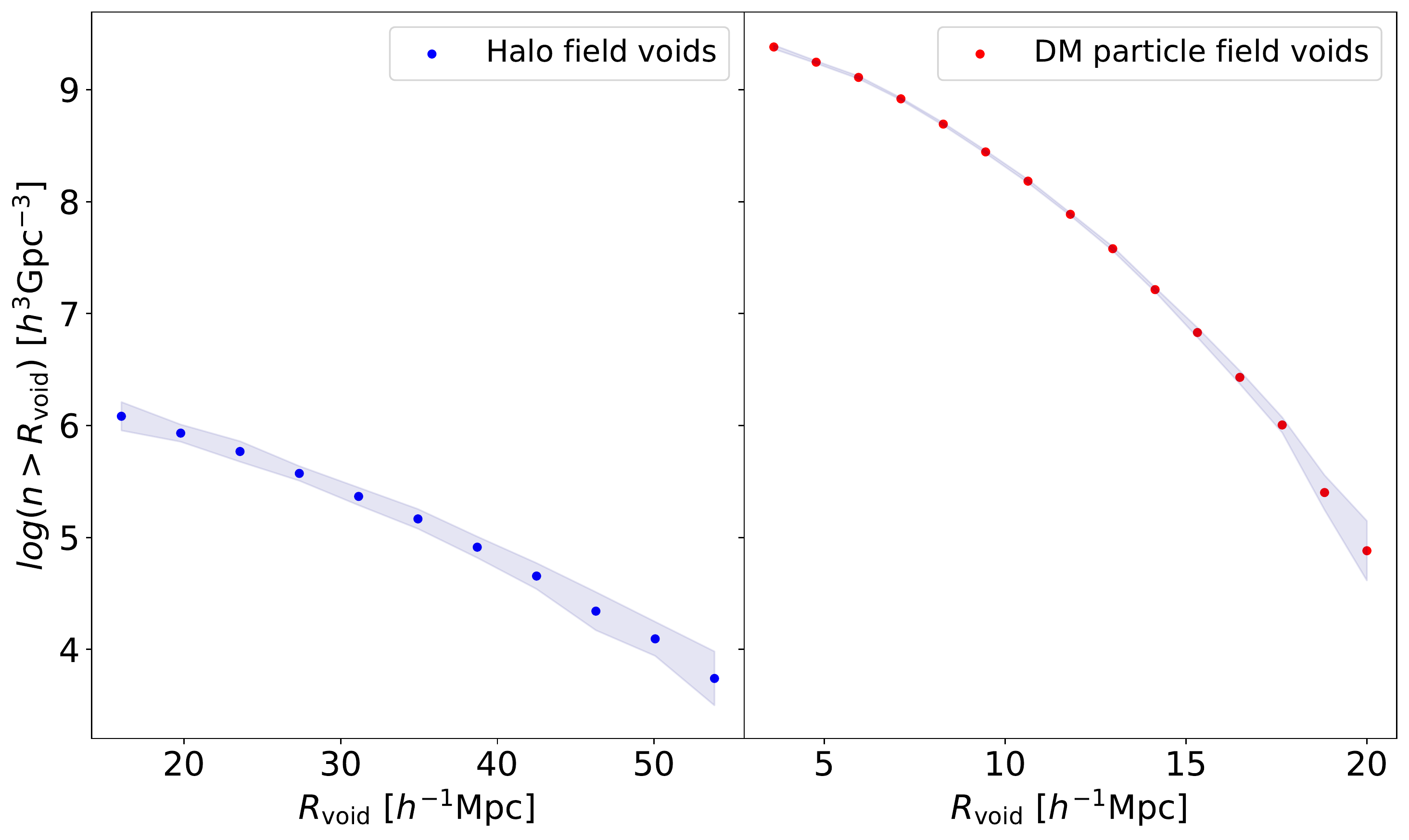}
    %\vspace*{-12mm}
    \caption{Cumulative void size function (number density of voids with radii above $R_{\rm void}$) for 21 realisations of \textsc{Gevolution} simulations at redshift $z=0.22$. The left figure illustrates the voids in the halo field, whereas the right figure shows the voids in the DM particle field. In each case, the shaded region represents the 1$\sigma$ error on the mean calculated from the 21 distinct realisations.}
    \label{fig3}
\end{figure*}
%%%%%%%%%%%%%%%%%%%%%%%%%%%%%%%%

%%%%%%%%%%% FIGURE %%%%%%%%%%%%%
\begin{figure*}
    \centering
    %\hspace*{-0.6in}
    \includegraphics[width=2\columnwidth]{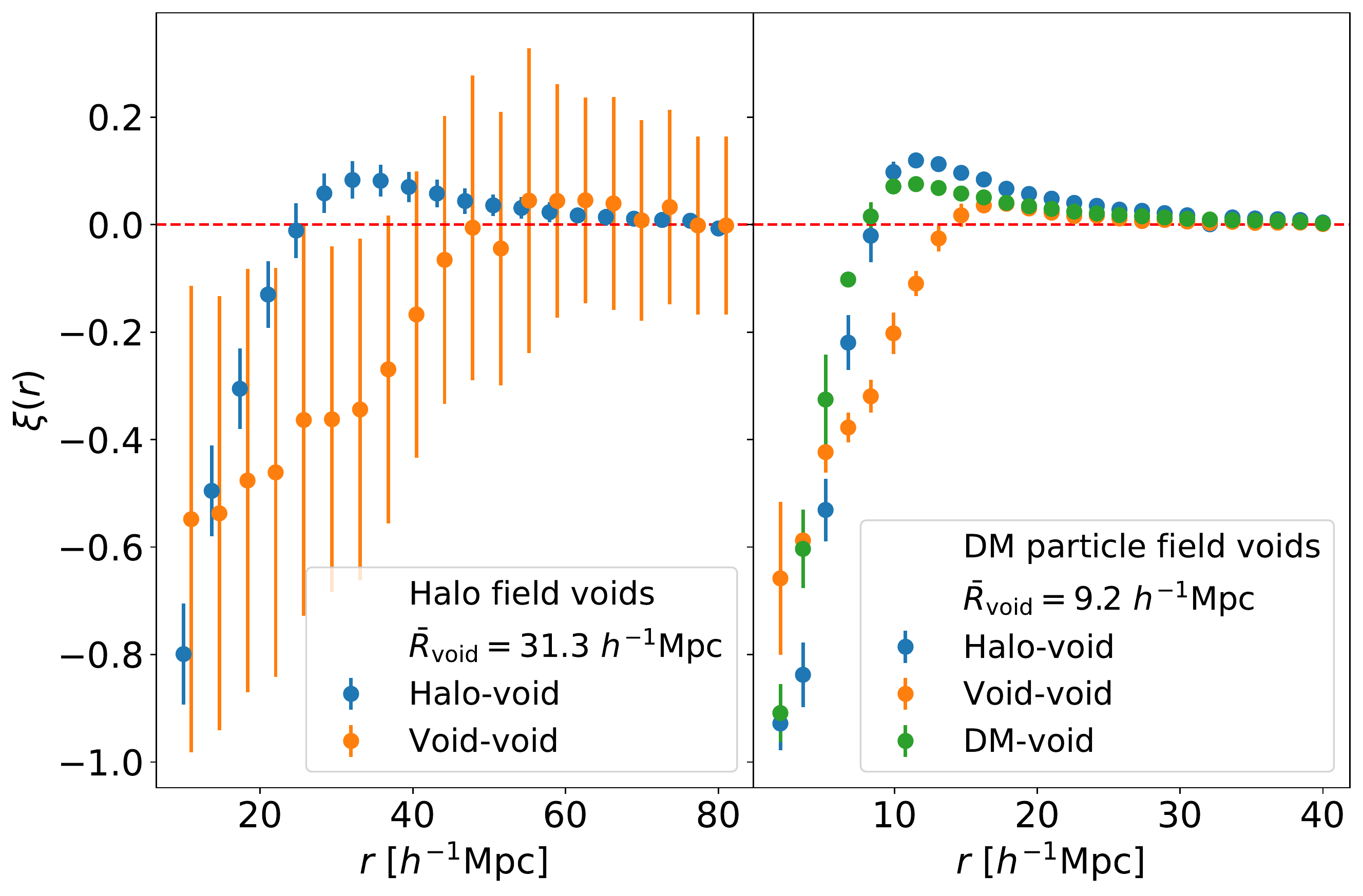}
    %\vspace*{-12mm}
    \caption{Variation of the halo-void, void-void, and DM-void 2PCF as a function of separation ($r$) for 21 realisations of \textsc{Gevolution} simulations at redshift $z=0.22$. The left figure illustrates the voids in the halo field, whereas the right figure shows the voids in the DM particle field. The error bars on both sides represent the 1$\sigma$ variation in the mean calculated from the 21 distinct realisations. Here all correlation functions are computed without applying any cut in radius. We artificially shift the void-void separation bins in the left side figure to improve the visual readability. }
    \label{fig5}
\end{figure*}
%%%%%%%%%%%%%%%%%%%%%%%%%%%%%%%%

\subsubsection{Doppler lensing algorithm}

Doppler lensing is a tool that can be used to measure the lensing signal (in conjunction with the weak-lensing signal) at low redshifts \citep{2013PhRvL.110b1302B, 2014MNRAS.443.1900B, 2017MNRAS.472.3936B,2022MNRAS.509.5142H}. Unlike weak gravitational lensing, which is caused by the mass distribution along the line of sight, Doppler lensing is purely related to the velocity of the source galaxies. Doppler lensing algorithm is based on solving Eqn. \ref{eqn:Dop} and the steps of computing Doppler convergence are as follows: 
\begin{itemize}
    \item \textit{Generating particle snapshots:} to generate particle snapshots at redshift $z=0.22$, we conduct 21 \textsc{Gevolution} simulations with different initial conditions but keeping the identical cosmological parameters. These are the same simulations and snapshots used in the weak-lensing algorithm part. 
    
    \item \textit{Finding cosmic structures:} by using \textsc{Rockstar}, we detect DM haloes and extract their information, such as positions, masses, velocities, and so on. Then, employing \textsc{Revolver}, we identify cosmic voids and extract their information, such as positions, radii, volume, and so on. 
    
    \item \textit{Targeting cosmic structures:} after identifying cosmic voids we subdivide them into three distinct radius ranges as like weak-lensing algorithm: i) $10~\hMpc \leq R_{\rm void} < 20~\hMpc$ ii) $20~\hMpc \leq R_{\rm void} < 30~\hMpc$, and iii) $30~\hMpc \leq R_{\rm void} < 40~\hMpc$.    
    
    \item \textit{Computing impact parameter:} then we compute the value of the impact parameter, which is the distance from the centre of the void to the centre of the haloes, modified by the factor of $\cos \theta$ where $\theta$ is the angle between the direction pointing from the observer towards the centre of the void, and the direction pointing from the centre of the voids towards the galaxy, cf. \citet{2022MNRAS.509.5142H}.
    
    \item \textit{Computing and stacking the Doppler convergence:} then we calculate the dot product of the normalised velocities ($\bm{v}/c$) of the haloes around cosmic voids and the direction vectors ($\bm{n}$) between the observer and the observed haloes. The Doppler convergence value is then obtained by multiplying the remaining parts of Eqn. \ref{eqn:Dop} with the dot product result. Finally, we stack Doppler convergence values for voids with three distinct subgroups of void radii.
\end{itemize}

\section{Void properties} \label{void_porperties}
\subsection{Void size function}

\citet{2004MNRAS.350..517S} have modelled for the first time the statistical analysis of the Void Size Function (VSF), also known as the void abundance, which is simply characterised as the number of voids in a given radius bin at a particular redshift. It is a potential approach for probing dynamical dark energy and modified gravity \citep{2012MNRAS.426..440B, 2015MNRAS.451.1036C, 2015PhRvD..92h3531P, 2019JCAP...12..040V}, constraining neutrino masses \citep{2015JCAP...11..018M, 2019MNRAS.488.4413K, 2019JCAP...12..055S, 2020PhRvD.102h3537Z, 2021MNRAS.504.5021C}, standard and coupled dark energy cosmologies \citep{2016MNRAS.455.3075P}, cubic Galileon and nonlocal gravity cosmologies \citep{2015JCAP...08..028B}, cluster-void degeneracy breaking \citep{2016ApJ...820L...7S}, etc.

Here we aim to focus on VSF for two different input tracers that are used to identify cosmic voids. Figure \ref{fig3} shows the cumulative VSF as a function of the effective void radius (number density of voids with radii above $R_{\rm void}$) for 21 realisations of \textsc{Gevolution} simulations at redshift $z=0.22$. As expected the VSF decreases with the increasing values of effective void radius in both halo field and DM particle field voids. It is observed from Fig. \ref{fig3} that the VSF is clearly dependent on the input tracers that are utilised to identify cosmic voids. In comparison to the voids identified in both fields, the DM particle field voids, for example, are smaller and more numerous than the halo field voids. The fundamental reason for such variability in void sizes and counts is that the distribution of haloes is sparser than the distribution of DM particles \citep{2014MNRAS.442..462S}.

\subsection{Void 2-point correlation function}

Here we aim to measure the 2-point correlation function (2PCF) for void-void, halo-void, and DM-void configurations. The void-galaxy cross-correlation function is a useful tool since it encodes the same information as the stacked radial density profile of voids (for more details see Section \ref{sec:density}). In recent decades, much emphasis has been paid to mapping the stacked profile of cosmic voids using the auto or cross-correlation function in both real and redshift space \citep{2013MNRAS.436.3480P, 2014PhRvL.112d1304H, 2015JCAP...11..036H, 2016MNRAS.459.4020L, 2016MNRAS.462.2465C, 2017MNRAS.469..787P, 2017PhRvD..95h3502A, 2018PhRvD..98d3527N, 2019MNRAS.483.3472N, 2019PhRvD.100b3504N, 2020ApJ...901...87P, 2022MNRAS.511.4333K}. We utilise the simple estimator proposed by \citet{1974ApJS...28...19P}, which has been implemented in the \textsc{Nbodykit} pipeline \citep{2018AJ....156..160H}, to compute the auto/cross-correlation function as
\begin{equation}\label{eq:natural_est}
    \xi(r)=\frac{DD(r)}{RR(r)}-1 ,
\end{equation}
where $D$ represents the data catalogue, $R$ is the synthetic random catalogue, $DD(r)$ and $RR(r)$ represent the pair counts with separation $r$ in the data and random catalogues respectively. It is worth noting that, in the context of uniform periodic randoms such as for periodic simulation boxes, \textsc{Nbodykit} analytically calculates the random pairings $RR(r)$ to minimise computing time.

The variation of the halo-void and void-void 2PCF as a function of separation ($r$) for 21 realisations of \textsc{Gevolution} simulations at redshift $z=0.22$ is shown in Fig. \ref{fig5}. These correlation functions are calculated using data from all voids, haloes, and DM particles (down-sampled) for each realisation of the simulation box at redshift $z=0.22$. Then for certain separation bins, we compute the mean auto/cross-correlation function and demonstrate it in Fig. \ref{fig5} as a function of the separation. As expected, the characteristics of these correlation functions are quite similar in both halo field and DM particle field voids. However, for halo field voids, higher separation bins are required because the sparser distribution of haloes results in larger void sizes than DM particle field voids. Again, the higher $1\sigma$ variation of the void-void correlation function for halo field voids is due to the smaller number of voids identified in the halo distribution than DM particle distribution. The profiles for these correlation functions are consistent with the existing literature \citep{2016MNRAS.459.4020L, 2016MNRAS.462.2465C, 2017MNRAS.469..787P, 2017PhRvD..95h3502A, 2018PhRvD..98d3527N, 2022MNRAS.511.4333K}.

For the void-void correlation function in both halo and DM particle fields, we observed that the peak of the void-void correlation function is approximately two times the mean effective void radius (i.e. $r \sim 2\bar{R}_{\rm void}$) of the sample, which is consistent with the existing literature \citep{2014JCAP...12..013H, 2021arXiv210702304K}. It is because void walls typically contact at twice the average effective void radius since they can't overlap, a consequence of void exclusion scale (for more details, see \citet{2014PhRvL.112d1304H, 2014PhRvD..90j3521C}). Due to this, the sample will likely have two void centres, each located at a distance equal to twice the sample's mean effective void radius \citep{2021arXiv210702304K}. Generally speaking, voids can't be separated by smaller sizes since they will inevitably merge. Additionally, void-void correlation function appears to have negligible systematic errors and biases from redshift-space distortions, making this statistic the cleanest one for identifying geometric distortions \citep{2014JCAP...12..013H}.

For the tracer-void (halo-void or DM-void) correlation function in both halo and DM particle fields, we observed that the peak of the tracer-void correlation function is approximately equal to the mean effective void radius (i.e. $r \sim \bar{R}_{\rm void}$) of the sample. The compensation walls of voids, which are composed of high-density structures including sheets, filaments, and clusters of galaxies, are the most dynamic areas. Galaxies far from the centre of the void have a net overdensity and so exhibit consistent infalls rather than outflows on large scales. So, random motions of these structures are responsible to observe the peak of the correlation function within the compensation wall of cosmic voids, a consequence of galaxy-void exclusion \citep{2005MNRAS.363..977P, 2014PhRvD..90j3521C}. It is also observed that the halo-void correlation function provides a slightly higher amplitude than the DM-void correlation function in the DM particle field voids (see right side of Fig. \ref{fig5}). %It could be the reason for the variability of DM particles in the halo and DM particle fields.
The reason for such variation is due to the tracer bias ($b_{\rm t}$) around cosmic voids. Since the distribution of haloes is more sparser than the distribution of DM particles (which indicates halo distributions exhibit stronger fluctuations than the DM), one may obtain a higher tracer bias factor ($b_{\rm t} > 1$) for computing the halo-void correlation function than the DM-void correlation function (for further information on how tracer bias impacts void properties, see these papers \citep{2017MNRAS.469..787P, 2019MNRAS.487.2836P}).

%%%%%%%%%%% FIGURE %%%%%%%%%%%%%
\begin{figure*}
    \centering
    %\hspace*{-0.6in}
    \includegraphics[width=2\columnwidth]{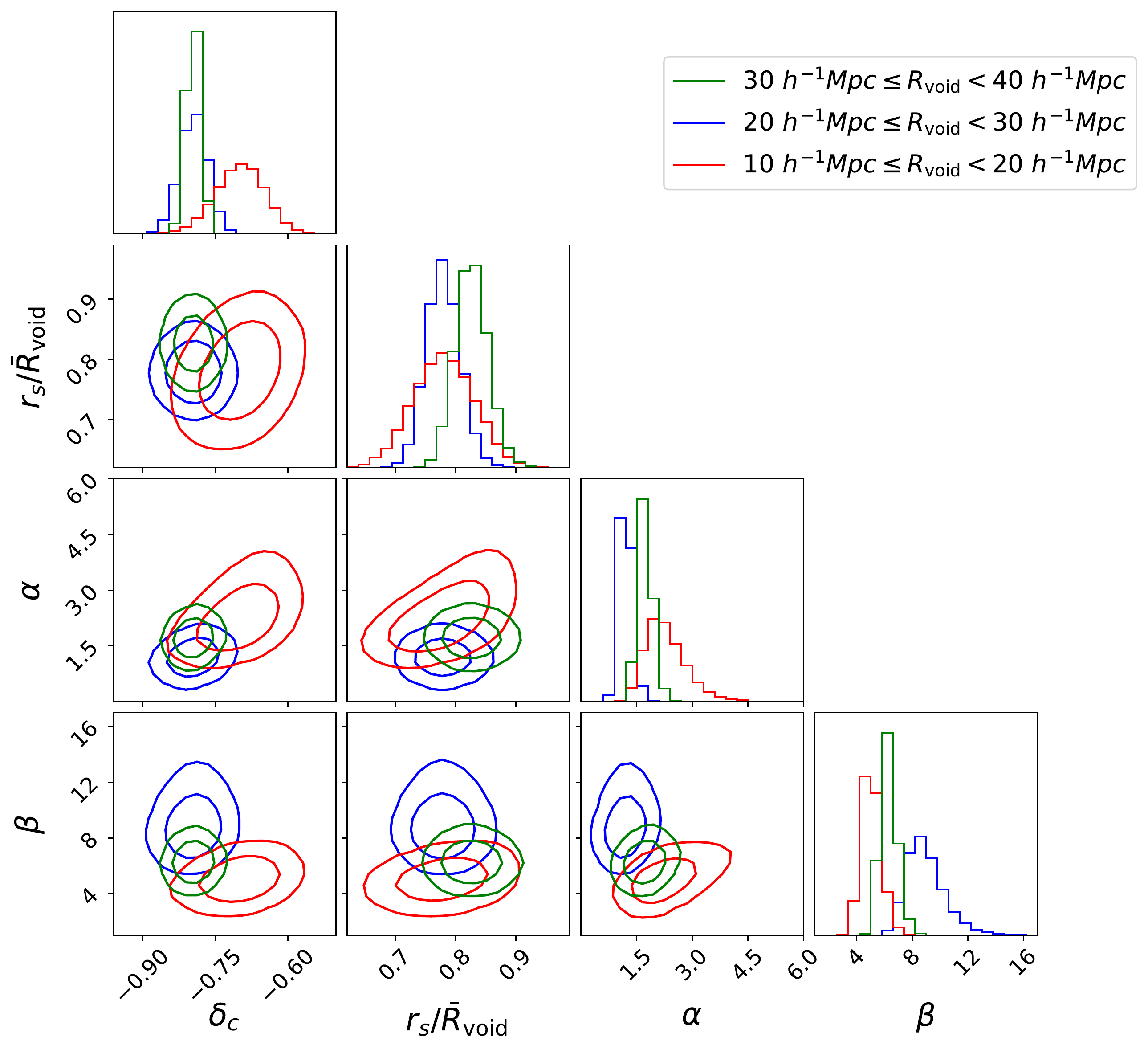}
    %\vspace*{-12mm}
    \caption{Posterior distribution of the model parameters of three distinct groups of void radii for 21 realisations of \textsc{Gevolution} simulations at redshift $z=0.22$. The 2D contours enclosing the 68\% and 95\% confidence levels, respectively. Parameter values at the maximum of the marginal posterior distribution are provided in table \ref{table:density_contour}. }
    \label{fig6a}
\end{figure*}
%%%%%%%%%%%%%%%%%%%%%%%%%%%%%%%%

\subsection{Void density profile}\label{sec:density}
Besides the void sizes, the distribution of the spherically averaged stacked matter density profile in and around cosmic voids is a significant quantity to shape our knowledge about voids \citep{2012ApJ...754..109L, 2012ApJ...761...44S, 2012ApJ...761..187S, 2013ApJ...762L..20K, 2014MNRAS.443.3238P, 2014MNRAS.440..601R, 2014PhRvL.112y1302H, 2015JCAP...03..047L, 2015MNRAS.449.3997N, 2015JCAP...08..028B, 2017MNRAS.469..787P, 2020ApJ...901...87P, 2021arXiv210913378R, 2021PhRvD.104b3512W}. \citet{2014PhRvL.112y1302H} introduced the following simple empirical function for capturing the spherically averaged stacked matter density profile in and around cosmic voids
\begin{equation}\label{eq:dens_prof}
  \frac{n_{\rm vt}}{\langle n_{\rm t} \rangle} -1 = \delta_c \frac{1-(r/r_s)^{\alpha}}{1+(r/R_{\rm void})^{\beta}}\,,
\end{equation}
where $\delta_c$ is the central density contrast at which the radial distance from the void centre $r=0$, $\langle n_{\rm t} \rangle$ is the average density of tracer, $r_s$ is a scale radius 
at which the void profile density $n_{\rm vt}$= $\langle n_{\rm t} \rangle$, and the inner and outer slopes of the void profile are represented by $\alpha$ and $\beta$.

%%%%%%%%%%% FIGURE %%%%%%%%%%%%%
\begin{figure}
  \centering
  \noindent
  \resizebox{\columnwidth}{!}{
  \includegraphics[width=\columnwidth]{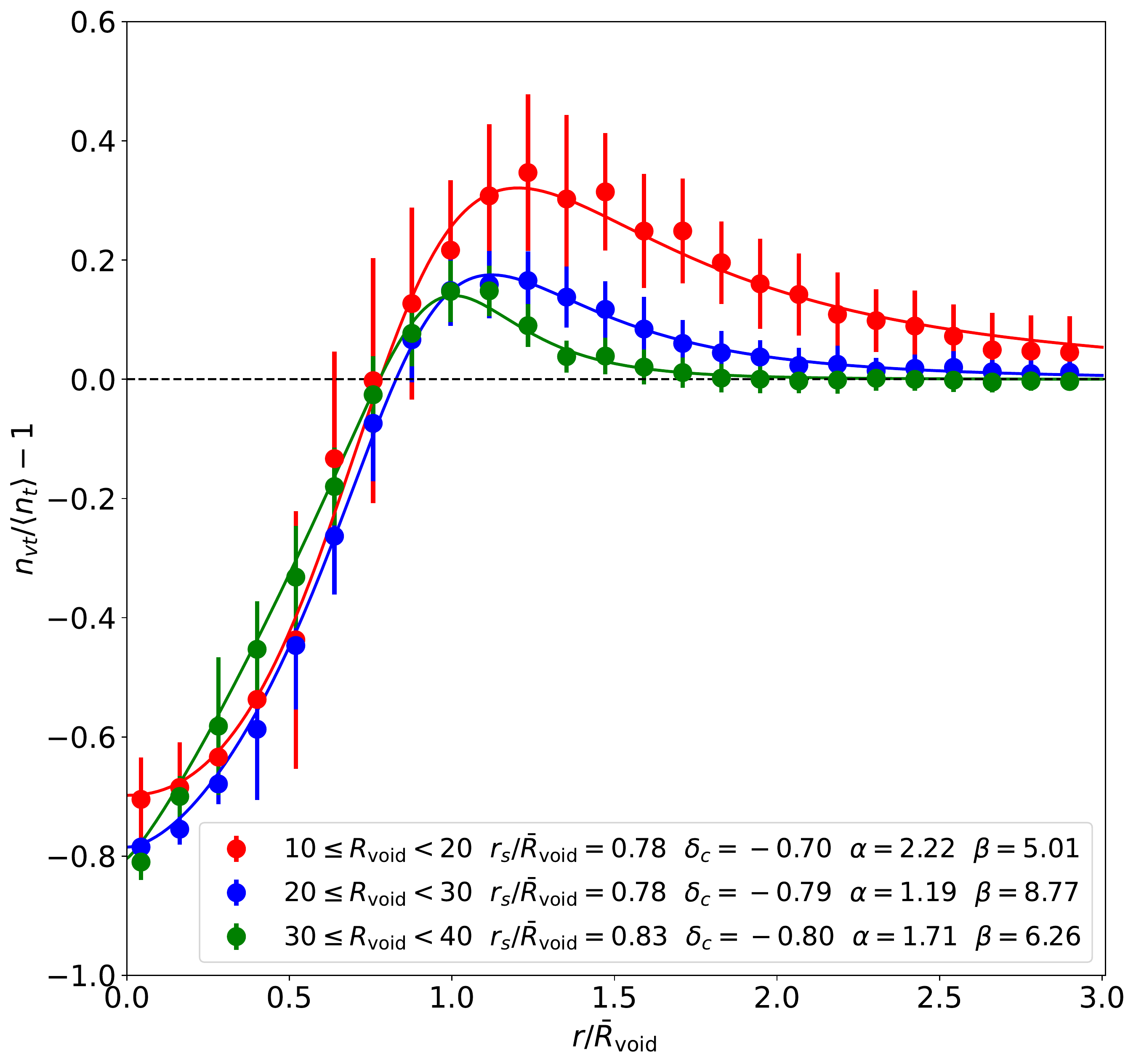}
  }
  \caption{The measured overdensity of tracer around tracer-void centres for three different subgroups of void radii as a function of impact parameter for 21 realisations of \textsc{Gevolution} simulations at redshift $z=0.22$. The solid lines represent the best fit of the tracer profile using Eqn. \ref{eq:dens_prof}. The error bars represent the 1$\sigma$ variation in the mean calculated from the 21 distinct realisations.}
  \label{fig6b}
\end{figure}
%%%%%%%%%%%%%%%%%%%%%%%%%%%%%%%%

Cross-correlation (void-tracers), by definition, counts tracers at a given distance from the void centre per unit volume, implying that the void-tracers correlation function carries the same information as the stacked radial density profile of voids \citep{2015JCAP...11..036H, 2017MNRAS.469..787P, 2019MNRAS.487.2836P}. Following \citet{2017MNRAS.469..787P} the average number density of the tracers from the centre of the cosmic voids, also known as the average stacked density profile of voids, can be mathematically expressed as 
\begin{align}
\dfrac{n_{\rm vt}(r)}{\langle n_{\rm t} \rangle} &= \dfrac{1}{N_{\rm v}} \sum_{i} \dfrac{n^i_{\rm vt}(r)}{\langle n_{\rm t} \rangle} \nonumber \\
&= \dfrac{1}{N_{\rm v}}\sum_i \dfrac{1}{N_{\rm t}} V\sum_j \delta^{D} (x_i^{\rm {c}} - x^{\rm t}_j +r) \nonumber \\
&= V \sum_{i,j} \int \dfrac{1}{N_{\rm v}} \delta^{D}(x_i^{\rm {c}}-x) \dfrac{1}{N_{\rm t}} \delta^{D}(x-x^{\rm t}_j +r) d^3 x \nonumber \\
&= \dfrac{1}{V} \int \dfrac{n_{\rm v}(x)}{\langle n_{\rm v} \rangle }  \dfrac{n_{\rm t}(x+r)}{\langle n_{\rm t} \rangle } d^3 x = 1+\xi_{\rm vt}(r),
\label{eq:stacked-void}
\end{align}
where r is the radial distance from the centre of the void, $N_{\rm v}$ is the number of voids, $N_{\rm t}$ is the number of tracers, $\langle n_{\rm v} \rangle$ is the average void density, $\langle n_{\rm t} \rangle$ is the average tracer density, $V$ is the total observed volume, $x_i^{\rm c}$ is the centre of the $i^{\rm th}$ void, $x_j^{\rm t}$ is the centre of the $j^{\rm th}$ tracer, $n_{\rm vt}$ is the void-tracer number density, $\delta^{D}$ is the Dirac delta function, and $\xi_{\rm vt}$ represents the void-tracer correlation function.

The identification of cosmic voids in the halo fields offers closer to real observational applications, so we are interested in the halo field voids from now. Another reason is that we will look at lensing measurements (which is related to the integrated stacked density profile of cosmic voids) in the context of ray-tracing algorithms later, and we will need to account for the complete picture of the density distribution of the simulation volume to capture the lensing characteristics adequately.

The halo-void correlation function (equivalent to the average density profile of cosmic voids) is then computed using \textsc{Nbodykit} for three distinct subgroups of void radii, and the theoretical model data is generated using Eqn. \ref{eq:dens_prof}. We then employ Markov Chain Monte Carlo (MCMC) method to estimate the posteriors of the model parameters from log-likelihood analysis. To perform this, we utilise
a publicly accessible code Emcee \citep{2013PASP..125..306F}. The following parameter vector summarises the free parameters of our model:
\begin{equation}
    \theta = (\delta_c, r_s, \alpha, \beta),
\end{equation}
and we consider the following uniform prior ranges for the values of our model parameters 
\begin{equation}
    \delta_c \in [-1,0], r_s \in [0.5,1.5]\Bar{R}_{\rm void}, \alpha \in [0.1,4.6], \beta \in [3,12]
\end{equation}

The results of the MCMC analysis in terms of the constraints on the parameters of the density profile are presented in Fig. \ref{fig6a}. The best fit models are  presented in Fig. \ref{fig6b}, which shows the variation of the average density profile of cosmic voids as a function of impact parameter ($r/\Bar{R}_{\rm void}$) for 21 realisations of \textsc{Gevolution} simulations at redshift $z=0.22$. We find good fits of our simulation data with model. The different colour markers show three distinct groups of void radii with best fit parameter values. It is clear from Fig. \ref{fig6b} in the regime $0 \leq r/\Bar{R}_{\rm void} < 1$, the profiles of average stacked density of cosmic voids for three distinct groups of void radii increase from minimum value ($\sim -0.8$) towards the background value i.e. $\sim 0$. The density profiles attain their maximal value in the regime $1 \leq r/\Bar{R}_{\rm void} < 2$, the region associated with the filaments and walls around the voids, and then start to drop with increasing the impact parameter. And, in the final regime $r/\Bar{R}_{\rm void} \geq 2$ the density distribution reaches its average value from the centre of the cosmic voids and the density profile of voids follows a constant background value, $\sim 0$. The properties of the stacked density profiles of voids presented here are consistent with existing literature \citep{2014MNRAS.440..601R, 2014PhRvL.112y1302H, 2015JCAP...03..047L, 2015MNRAS.449.3997N, 2015JCAP...08..028B, 2017MNRAS.469..787P, 2020ApJ...901...87P, 2021arXiv210913378R, 2021PhRvD.104b3512W}.

\begin{table*}
\begin{minipage}{100mm}
\caption{The best fit parameter values for computing stacked average density profile of cosmic voids for three distinct groups of void radii.}
\begin{tabular}{@{}cccccccc}
\hline
Void radius [$h^{-1}\mathrm{Mpc}$]  & $\delta_c$ & $r_s/\Bar{R}_{\rm void}$
& $\alpha$  & $\beta$  &\\
\hline
$30 \leq R_{\rm void} < 40$ & $-0.80 \pm 0.01$ & $0.83 \pm 0.03$ & $1.71 \pm 0.16$ & $6.26 \pm 0.63$ \\
$20 \leq R_{\rm void} < 30$  & $-0.79 \pm 0.03$ & $0.78 \pm 0.03$ & $1.19 \pm 0.17$ & $8.79 \pm 1.49$  \\
$10 \leq R_{\rm void} < 20$ & $-0.70 \pm 0.05$ & $0.78 \pm 0.05$ & $2.22 \pm 0.61$ & $5.01 \pm 0.71$ \\
\hline\\
\end{tabular}
\label{table:density_contour}
\end{minipage}
\end{table*}

\section{Lensing Measurements} \label{lensing_measurements}
\subsection{Model fits} \label{lensing_theory}
Integrating the stacked density profile as described in Eqn. \ref{eq:dens_prof} along the line of sight provides the theoretical prediction for the statistical lensing measurement. The projected surface mass density and differential surface mass density are connected to the lensing statistical quantities (i.e. convergence, shear, and magnification). Following \citet{2013ApJ...762L..20K}, the differential surface mass density can be expressed as  
\begin{equation}
    \Delta\Sigma(r/R_{\rm void}) = \bar{\Sigma}(<r/R_{\rm void}) - \Sigma(r/R_{\rm void}) \, ,
\end{equation}
where the projected surface mass density can be written as
\begin{equation}
    \Sigma(r/R_{\rm void}) = \int {\rm d} r_{\rm los} {} \, \left ( n_{\rm v} (r) - \bar{n} \right )\, ,
\end{equation}
where 
\begin{equation}
r = \sqrt{ r_{\rm los}^2  + d_{\rm void}^2  - 2 \, r_{\rm los} \, d_{\rm void} \cos \varphi }, 
\end{equation}
and  $r_{\rm los}$ is the line-of-sight distance,
$d_{\rm void}$ is the distance to the centre of the void, and 
$ \varphi$ is the angle between the direction to the centre of the voids and direction of the line of sight; finally   
 $\bar{n}$ is the cosmological mean mass density. Therefore, the theoretical predictions for weak-lensing convergence ($\kappa_{\rm theory}$) and magnification ($\mu_{\rm theory}$) can be written as
\begin{equation} \label{eq:convergence}
    \kappa_{\rm theory} = \frac{1} {\Sigma_{\rm cr}} 
    \int {\rm d} r_{\rm los} \left( n_{\rm v} - \bar n \right)  , 
\end{equation}
\begin{equation} \label{eq:magni}
    \mu_{\rm theory} \approx 1 + \frac{2} {\Sigma_{\rm cr}} 
    \int {\rm d} r_{\rm los} \left( n_{\rm v} - \bar n \right)  ,  
\end{equation}
where $n_{\rm los}$ is the density along the line-of-sight and can be represented as Eqn. \ref{eq:dens_prof}. And, the weak-lensing tangential shear profile can be expressed as 
\begin{equation} \label{eq:shear}
    \gamma_{\rm theory} = \bar{\kappa}_{\rm theory}(<r/R_{\rm void}) - \kappa_{\rm theory}(r/R_{\rm void}) \, .
\end{equation}

\subsection{MCMC analysis and likelihoods}

We use a MCMC approach to estimate parameter posteriors and generate likelihood contours for the entire parameter space. We employ consistent priors, as described in Section \ref{sec:density} but for lensing measurement we fix $\alpha$ and $\beta$ parameters (by taking the best fit parameter values listed in Table \ref{table:density_contour}) and estimate the model parameters using log-likelihood analysis. We compute the likelihood functions for measuring weak-lensing shear and magnification, as well as Doppler convergence, using the corresponding theoretical predictions (as described in Section \ref{dop} and Section \ref{lensing_theory}) and numerical data, but for combined lensing analysis, we compute the total log-likelihood as the sum of all log-likelihood functions as
\begin{equation}\label{eq:log-likelihoods} 
    \log {\cal L}_{\rm total}  =  \log ( {\cal L}_{\mathrm{WL}-\mu}  ) + \log ( {\cal L}_{\mathrm{WL}-\gamma} ) + \log ( {\cal L}_{\mathrm{DL}- \kappa_{\nu}}) ,
\end{equation}
where ${\cal L}_{\mathrm{WL}-\mu}$, ${\cal L}_{\mathrm{WL}-\gamma}$, and ${\cal L}_{\mathrm{DL}-\kappa_{\nu}}$ are the weak-lensing magnification, weak-lensing shear, and Doppler lensing convergence likelihoods, respectively. The log-likelihood of the weak-lensing shear and magnification, and Doppler convergence can be expressed as 
\begin{equation}\label{eq:log-likelihood} 
    \log ( {\cal L}_{s}  ) = - \frac{1}{2}
    \sum \left[ \left( \frac{s_{\rm data} - s_{\rm theory} } { \sigma_{\rm data}} \right)^2 + \log(2\pi \sigma_{\rm data}^2)\right],
\end{equation}
where $s$ can be weak-lensing shear ($\gamma$), magnification ($\mu$), and Doppler convergence ($\kappa_{\nu}$) and $\sigma_{\rm data}$ is the error of the numerical data of each bin.

\begin{figure}
  \centering
  \noindent
  \resizebox{\columnwidth}{!}{
  \includegraphics[width=\columnwidth]{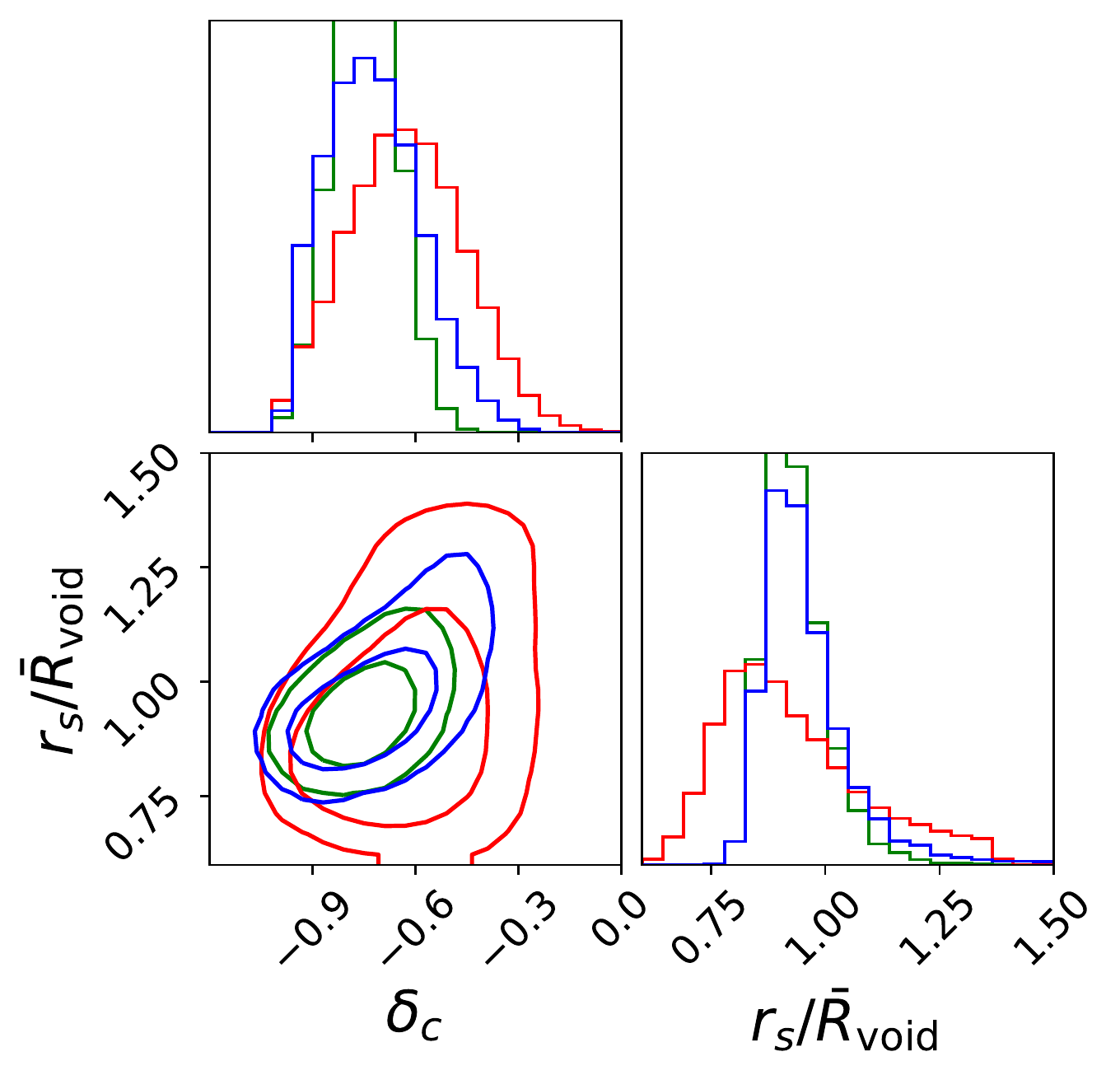}
  }
  \caption{Posterior distribution of the model parameters for weak-lensing shear measurement by stacking voids having three distinct groups of void radii at lens redshift $z=0.22$. The green, blue, and red lines represent voids having radii $30\hMpc<R_{\rm void}<40\hMpc$, $20\hMpc<R_{\rm void}<30\hMpc$, $10\hMpc<R_{\rm void}<20\hMpc$, respectively. The 2D contours enclosing the 68\% and 95\% confidence levels, respectively.}
  \label{fig8a}
\end{figure}

We first use the MCMC approach to constrain the parameters of a void using the gravitational lensing shear. 
%We keep the model parameters and the ranges of the priors as described in Section \ref{sec:density} and use the log-likelihood analysis to estimate the model parameters. 
Figure \ref{fig8a} shows the posterior distribution of the model parameters for weak-lensing tangential shear measurement by stacking voids having three distinct groups of void radii, and table \ref{table:lensing_contour} lists the best fit parameter values for computing stacked weak-lensing tangential shear around cosmic voids. It is clear from Fig. \ref{fig8a} that the bigger voids provide the smallest contours for weak-lensing shear measurements, whereas the smaller voids give the largest contours, which is consistent with the density contrast contours of cosmic voids in Fig. \ref{fig6a}. It is likely because the average number density in cosmic voids is higher for smaller voids, causing fluctuations in the number density of void galaxies inside cosmic voids, leading to higher statistical uncertainties and weakening the constraining power as compared to bigger voids. 
%The characteristical result of weak-lensing tangential shear is consistent with the existing literature \citep{2013ApJ...762L..20K, 2014MNRAS.440.2922M, 2015MNRAS.454.3357C, 2017MNRAS.465..746S}.
However, it is worth noting that due to the effects of tracer bias around cosmic voids \citep{2017MNRAS.469..787P, 2019MNRAS.487.2836P}, the shear constraints underestimate density $\delta_c$ and overestimate the parameter $r_s$ as compared to halo tracers.

%%%%%%%%%%% FIGURE %%%%%%%%%%%%%
%%%%%%%%%%%%%%%%%%%%%%%%%%%%%%%%

%%%%%%%%%%% FIGURE %%%%%%%%%%%%%
\begin{figure}
  \centering
  \noindent
  \resizebox{\columnwidth}{!}{
  \includegraphics[width=\columnwidth]{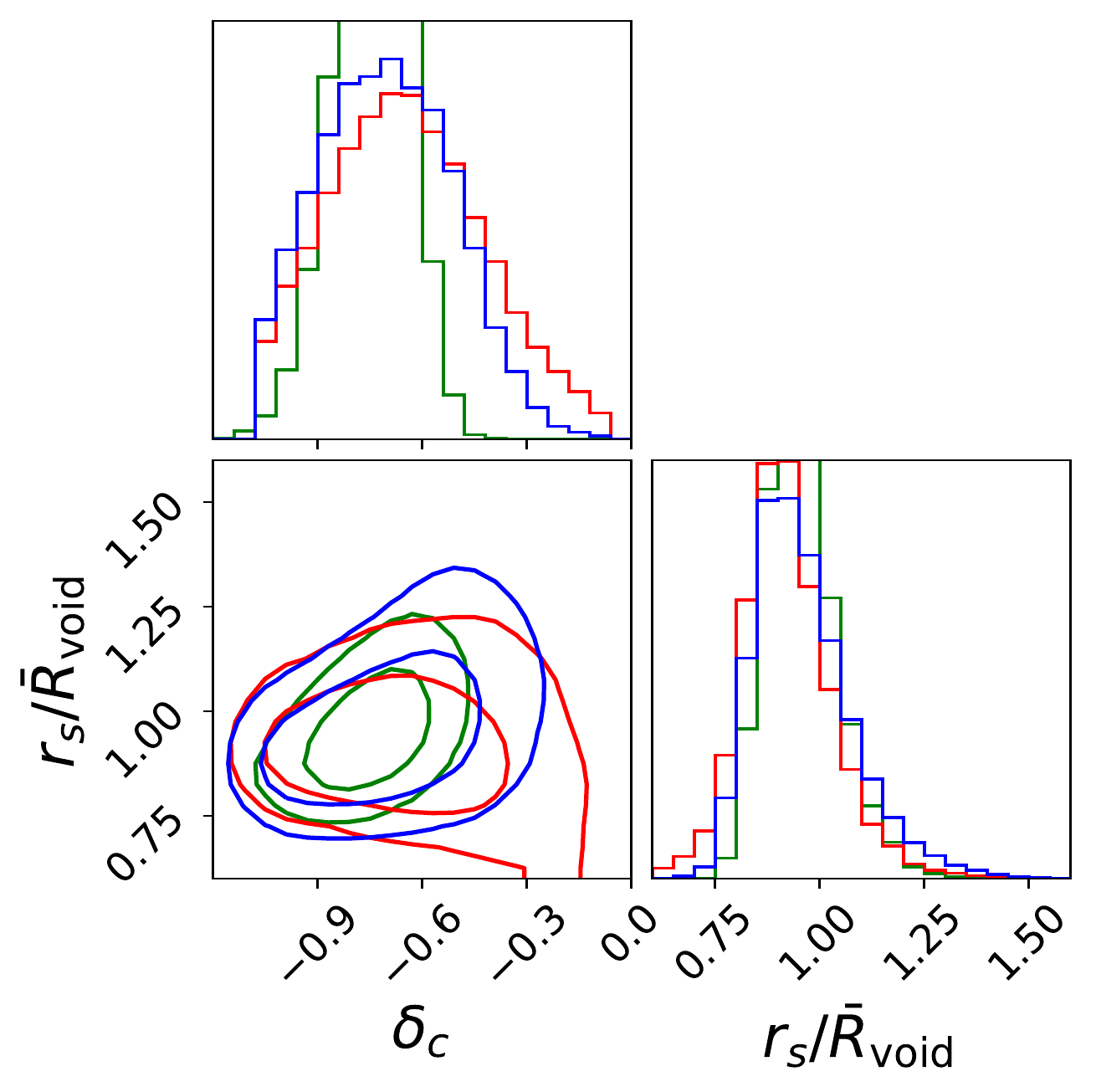}
  }
  \caption{The same as Fig. \ref{fig8a} but for the weak-lensing magnification measurement.
  %Posterior distribution of the model parameters for weak-lensing magnification measurement by stacking voids having three distinct groups of void radii at lens redshift $z=0.22$. The blue, orange, and green lines represent voids having radii $30\hMpc<R_{\rm void}<40\hMpc$, $20\hMpc<R_{\rm void}<30\hMpc$, $10\hMpc<R_{\rm void}<20\hMpc$, respectively. The 2D contours enclosing the 68\% and 95\% confidence levels, respectively.
  }
  \label{fig11}
\end{figure}
%%%%%%%%%%%%%%%%%%%%%%%%%%%%%%%%

%%%%%%%%%%% FIGURE %%%%%%%%%%%%%
\begin{figure}
  \centering
  \noindent
  \resizebox{\columnwidth}{!}{
  \includegraphics[width=\columnwidth]{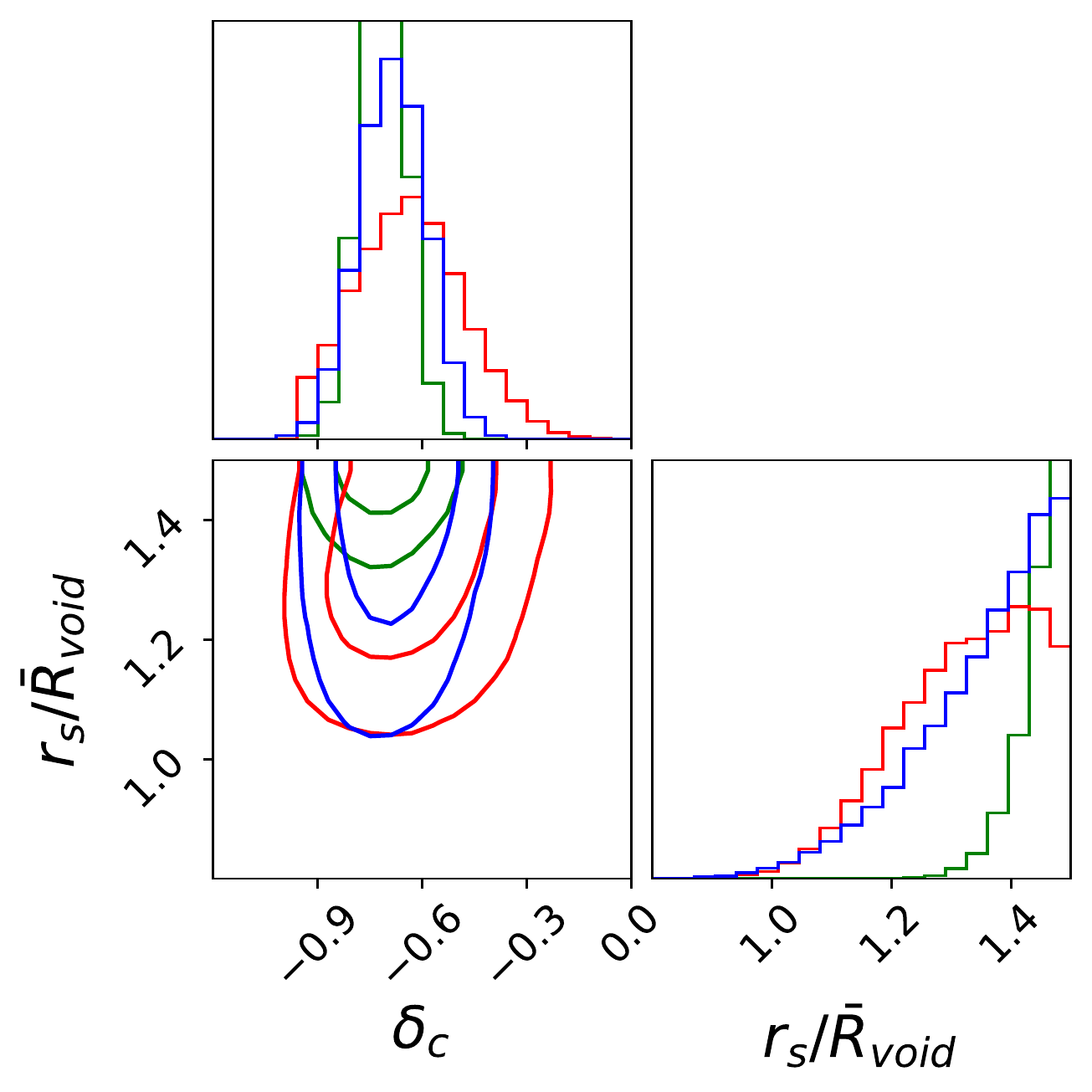}
  }
  \caption{The same as Fig. \ref{fig8a} but for the Doppler lensing convergence measurement.
  %Posterior distribution of the model parameters for weak-lensing magnification measurement by stacking voids having three distinct groups of void radii at lens redshift $z=0.22$. The blue, orange, and green lines represent voids having radii $30\hMpc<R_{\rm void}<40\hMpc$, $20\hMpc<R_{\rm void}<30\hMpc$, $10\hMpc<R_{\rm void}<20\hMpc$, respectively. The 2D contours enclosing the 68\% and 95\% confidence levels, respectively.
  }
  \label{fig_dop}
\end{figure}
%%%%%%%%%%%%%%%%%%%%%%%%%%%%%%%%

We next investigate the constraints from weak-lensing magnification. The contours of the model parameters for weak-lensing magnification measurement for three distinct groups of void radii is depicted in Fig. \ref{fig11}. The contour plots generated from the weak-lensing magnification measurement, like the weak-lensing tangential shear measurement, are the largest for smaller voids as compared to the bigger ones. But the constraints of the lensing model parameters are tighter for the weak-lensing tangential shear measurement than those for the weak-lensing magnification. 
%The theoretically best-fit profile of weak-lensing magnification is illustrated in Fig. \ref{fig11} by the solid lines. 
%The typical outcome of the weak-lensing magnification profile is consistent with the existing literature \citep{2013ApJ...762L..20K, 2013MNRAS.432.1021H, 2018MNRAS.480L.101D}.
As in the case of the weak-lensing shear, when compared against the profile 
obtained based on the halo tracers, the weak-lensing magnification underestimates density $\delta_c$ and overestimates $r_s$. Due to the fluctuation of the density distribution, voids identified in the halo field provide a higher tracer bias factor than the DM particle field \citep{2017MNRAS.469..787P, 2019MNRAS.487.2836P}. This tracer bias around cosmic voids is responsible for the underestimation of density $\delta_c$ when a halo tracer is used rather than a DM tracer. Figure \ref{fig_dop} shows the contours of the model parameters for Doppler convergence measurement for three distinct groups of void radii. We observed that $\delta_c$ looks good but $r_s$ is not constrained by Doppler lensing.

\begin{figure}
  \centering
  \noindent
  \resizebox{\columnwidth}{!}{
  \includegraphics[width=\columnwidth]{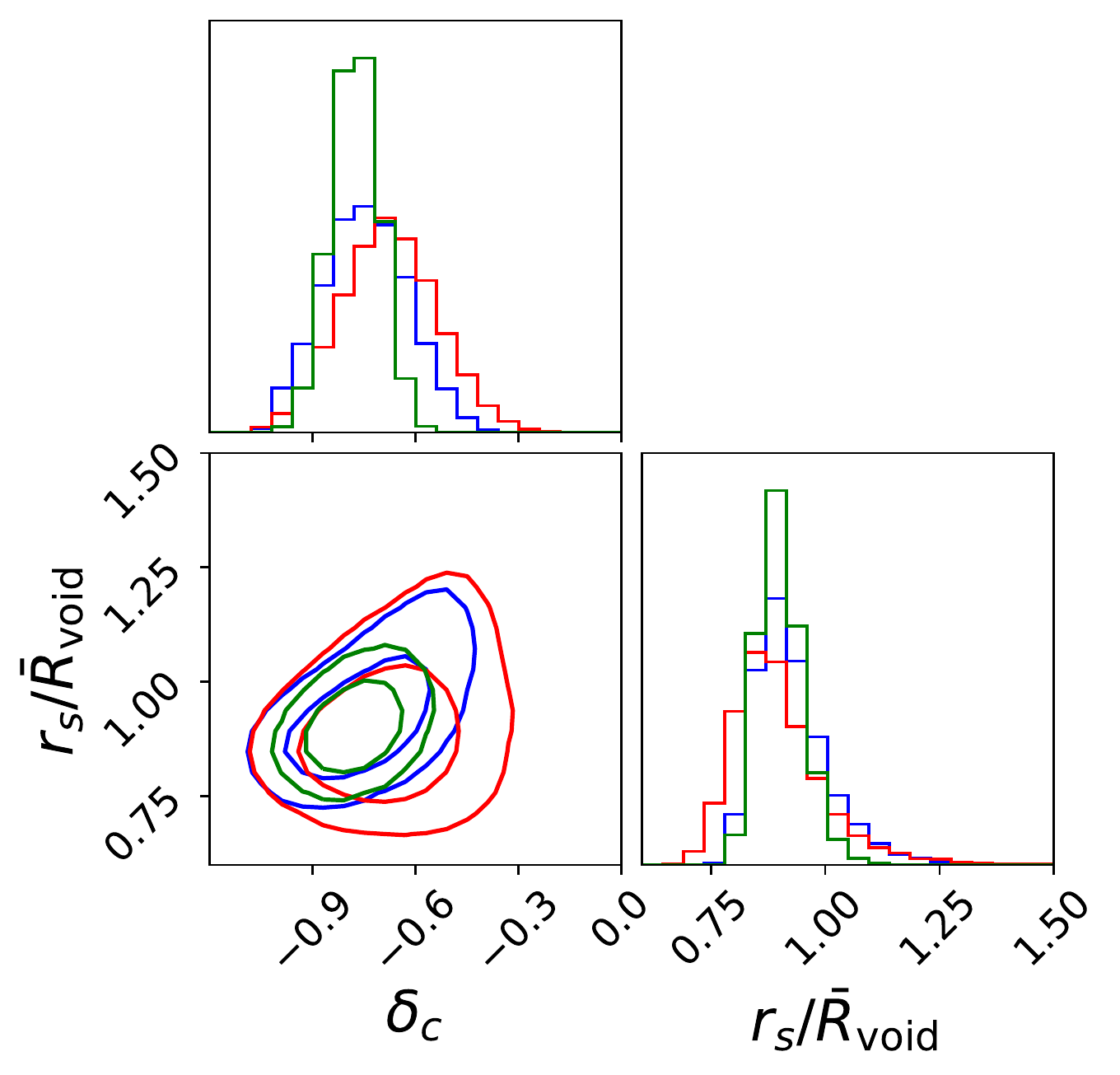}
  }
  \caption{The same as Fig. \ref{fig8a} but for the combined lensing (weak+Doppler lensing) measurement.
  %Posterior distribution of the model parameters for combined lensing (weak+Doppler lensing) measurement by stacking voids having three distinct groups of void radii at lens redshift $z=0.22$. The blue, orange, and green lines represent voids having radii $30\hMpc<R_{\rm void}<40\hMpc$, $20\hMpc<R_{\rm void}<30\hMpc$, $10\hMpc<R_{\rm void}<20\hMpc$, respectively. The 2D contours enclosing the 68\% and 95\% confidence levels, respectively.
  }
  \label{fig12}
\end{figure}

\begin{figure}
  \centering
  \includegraphics[width=\columnwidth]{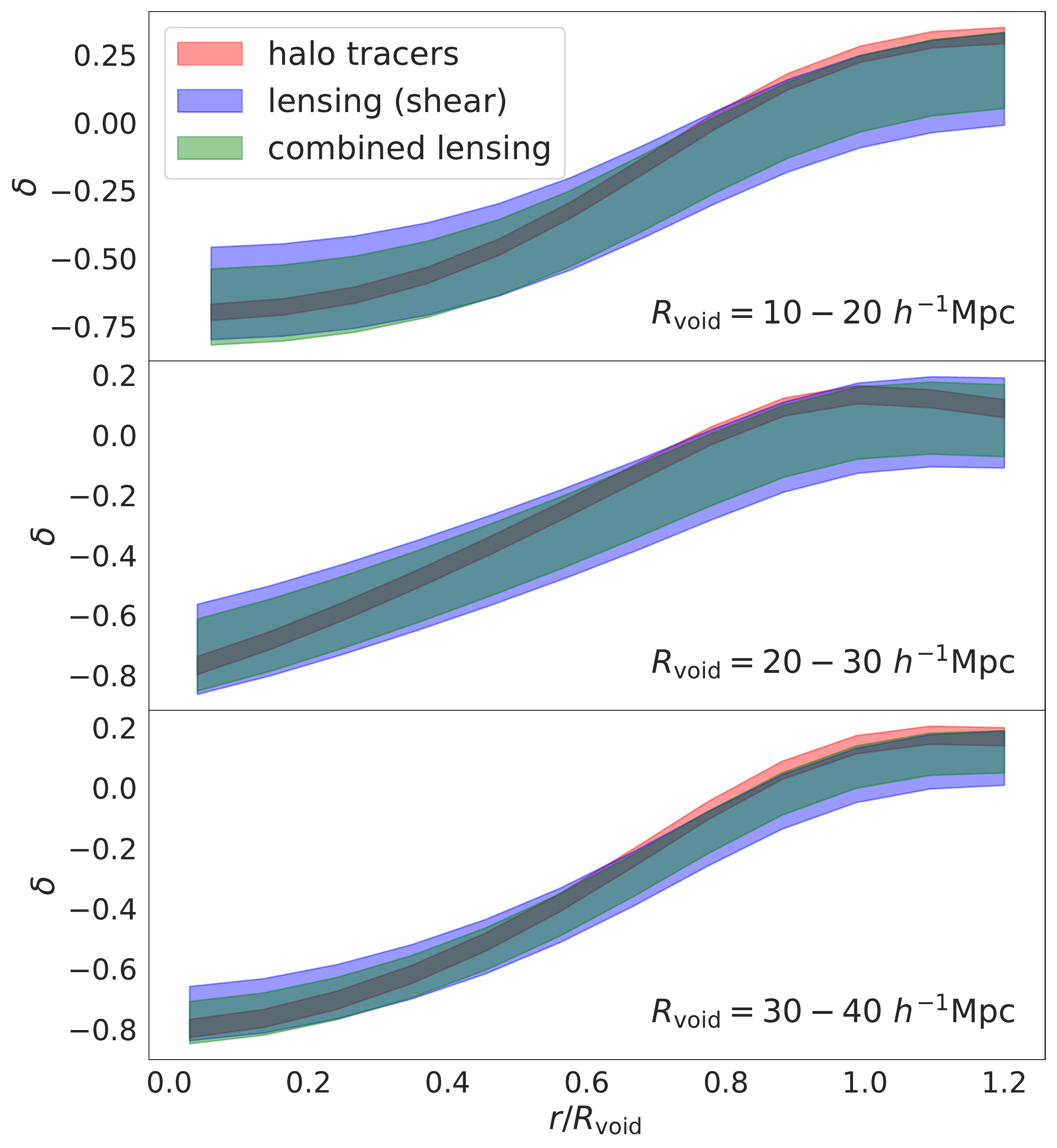}
  \caption{Density profiles of cosmic voids based on the halo tracers (red), weak-lensing shear (blue), and combined lensing (i.e. weak-lensing + Doppler lensing) (green).} 
  \label{fig-density}
\end{figure}

Finally, we want to combine all the lensing signals (weak-lensing shear + weak-lensing magnification + Doppler lensing). Figure \ref{fig12} shows the posterior distribution of the model parameters for combined lensing measurement by stacking voids having three distinct groups of void radii at lens redshift $z_l = 0.22$. Table \ref{table:lensing_contour} lists the best fit parameter values for combined lensing analysis. For combined lensing analysis, we found a better improvement of the contour size for all ranges of void radius, and the size of the contours is larger for smaller voids as compared to bigger ones.  It is apparent from Figs. \ref{fig12} and 
\ref{fig-density} the constraints of the lensing model parameters are significantly tighter for the combined lensing case than weak-lensing alone.

% %%%%%%%%%%% FIGURE %%%%%%%%%%%%%
% %%%%%%%%%%%%%%%%%%%%%%%%%%%%%%%%

%[scale=0.67]
% %%%%%%%%%%% FIGURE %%%%%%%%%%%%%
% %%%%%%%%%%%%%%%%%%%%%%%%%%%%%%%%

%%%%%%%%%%% TABLE %%%%%%%%%%%%%
\begin{table}
\begin{center}
\begin{tabular}{@{}llcccc@{}}
\toprule
  & Statistics                        &
  \multicolumn{1}{l}{Void radius [$h^{-1}\mathrm{Mpc}$]} &\multicolumn{1}{l}{$\delta_c$} & \multicolumn{1}{l}{$r_s/\Bar{R}_{\rm void}$} & \\ \midrule
& $\gamma$ & & $-0.75 \pm 0.09$ & $0.93 \pm 0.07$         \\

& $\mu$ & $30 \leq R_{\rm void} < 40$       & $-0.74 \pm 0.10$                                & $0.95 \pm 0.09$                     \\

& $\kappa_{\nu}$ &   & $-0.72 \pm 0.06$                                & $1.46 \pm 0.04$                      \\

& Combined &   & $-0.78 \pm 0.07$                                & $0.90 \pm 0.05$         
%\bottomrule
\\ \midrule
& $\gamma$ &     & $-0.73 \pm 0.15$                              & $0.94 \pm 0.10$         \\

& $\mu$ & $20 \leq R_{\rm void} < 30$       & $-0.71 \pm 0.20$                                & $0.94 \pm 0.11$                     \\

& $\kappa_{\nu}$ &   & $-0.70 \pm 0.10$                                & $1.38 \pm 0.11$                      \\

& Combined &   & $-0.75 \pm 0.12$                                & $0.91 \pm 0.09$         
%\bottomrule
\\ \midrule
& $\gamma$ &  & $-0.63 \pm 0.17$                              & $0.90 \pm 0.17$         \\

& $\mu$ & $10 \leq R_{\rm void} < 20$       & $-0.66 \pm 0.24$                                & $0.91 \pm 0.11$                     \\

& $\kappa_{\nu}$ &   & $-0.64 \pm 0.15$                                & $1.34 \pm 0.14$                      \\

& Combined &  & $-0.68 \pm 0.14$      & $0.88 \pm 0.10$         \\ 
\bottomrule

\end{tabular}
\caption{ The best fit parameter values for computing stacked lensing statistical quantities around the cosmic voids for three distinct groups of void radii.}
\label{table:lensing_contour}
\end{center}
\end{table}
%%%%%%%%%%%%%%%%%%%%%%%%%%%%%%%%

\section{Conclusion and Outlook} \label{conclusions}

Weak gravitational lensing and Doppler lensing signals have been examined numerically in this work within the framework of relativistic $N$-body simulations. We began by examining the different features of cosmic voids, such as the VSF, the void 2PCF, and the void density profile, for two distinct sets of input tracers. Numerous studies have identified these void characteristics extensively, but this is the first time we have compared the findings for halo-field and DM particle field voids in the context of relativistic $N$-body simulations. Then, using our ray-tracing technique (3D-RBT), we projected bundles of photons in the direction of cosmic voids with three distinct groups of void radii and extracted the statistical quantities associated with weak-lensing, namely convergence, shear, and magnification. We also constructed a 3D Doppler lensing algorithm to quantify the Doppler convergence around the cosmic voids. We investigated the impact of mass distribution on lensing signals around cosmic voids and ran MCMC samplings from our mock weak-lensing and Doppler lensing data to forecast the accuracy for measuring lensing signals.

% VSF
For the analysis of cosmic void properties, we considered two tracer fields (such as DM particles and haloes) and then identified cosmic voids from those tracer fields. Due to the sparser halo distributions, the number of cosmic voids in the DM particle field is smaller and more numerous than those in the halo field. To begin, we examined the VSF characteristics as a function of the radius of cosmic voids and observed that they are highly dependent on the tracer used to identify cosmic voids (cf. Fig. \ref{fig3}). 
%mass
%The mass profile of cosmic voids also significantly depends on the input tracers. We showed that the bigger voids are more massive as compared to smaller ones (cf. Fig. \ref{fig4}).
% 2PCF
The characteristics of halo-void and void-void 2PCF are quite similar in both halo field and DM particle field voids (cf. Fig. \ref{fig5}). The void-void correlation function for halo field voids has a higher statistical uncertainty due to the smaller number of voids identified in the halo distribution than DM particle distribution. 
% density contrast
Then we analysed the density contrast of cosmic voids as a function of impact parameter and showed that the characteristic profiles of density contrast for three distinct groups of void radii (cf. Fig. \ref{fig6b}) follow the universal trends \citep{, 2013ApJ...762L..20K, 2014MNRAS.440..601R, 2014PhRvL.112y1302H, 2015MNRAS.449.3997N, 2015JCAP...08..028B}.

The most important result of this paper was the analysis of constraining lensing model parameters from the theoretical predictions and numerical data, followed by the combination of weak-lensing and Doppler lensing signals to increase constraining power. The weak-lensing tangential shear profile provides comparatively tighter constraints for the density contrast as compared to the weak-lensing magnification for all three distinct groups of void radii. The combined lensing (weak-lensing shear + weak-lensing magnification + Doppler lensing convergence) analysis generated contours are smaller than any of the individual contours (e.g. weak-lensing shear or magnification), for all combinations of parameters. It shows that the combined weak-lensing and Doppler lensing signals improved the constraints, and also brings it to a better agreement with the profile extracted from the halo-tracers. 
%As shown in  Fig. \ref{fig-density}, this works well for smaller voids, however, for large voids, there is an apparent disagreement between the methods based on lensing and halo-tracers. 
The results of this study should be considered as a proof-of-concept that studied and demonstrated the capability of mapping the matter distribution inside cosmic voids using Doppler lensing in combination with gravitational lensing. The findings provided here are applicable to current and future low-redshift spectroscopic surveys, such as the DESI Bright Galaxy survey \citep{2020RNAAS...4..187R, 2022MNRAS.509.1478Z}, the 4MOST survey \citep{2019Msngr.175....3D}, etc.

%applications

Recently, \citet{2021MNRAS.507.2267D} have shown that combining void abundance with weak-lensing shear substantially enhanced the constraining power and resulted in tighter constraints on the measurement of cosmological parameters. It would be interesting to assess the constraining power by combining void abundance, weak-lensing statistics, and Doppler lensing signal to precisely measure the lensing model parameters and constrain the cosmological parameters. The covariance effects on the joint measurement of the weak-lensing and Doppler lensing signals and taking shape noise into account would be interesting, but we leave this for future studies.

\section*{Acknowledgements}

We acknowledge the use of Artemis at The University of Sydney for providing HPC resources that have contributed to the research results reported within this paper. We thank the anonymous referee for providing useful remarks that contributed to the final form of this paper.
MRH and SAE are supported by the Australian Government and the University of Sydney through the Research Training Program (RTP) Scholarship.
MRH would like to thank Florian List and William H. Oliver for their discussions during this work. 
KB acknowledges support from the Australian Research Council through the Future Fellowship FT140101270. This research work made use of the free Python packages \textsc{Numpy}  \citep{2020Natur.585..357H}, \textsc{Matplotlib} \citep{2007CSE.....9...90H}, \textsc{Scipy} \citep{2020SciPy-NMeth}, \textsc{Emcee} \citep{2013PASP..125..306F}, \textsc{Corner} \citep{2016JOSS....1...24F}, \textsc{H5py}\footnote{{\textcolor{blue}{https://www.h5py.org/}}},  \textsc{Eqtools}\footnote{{\textcolor{blue}{https://eqtools.readthedocs.io/en/latest/}}}, and \textsc{Mpi4py} \citep{DALCIN20051108}.

\section*{Data Availability}
The data generated as part of this project may be shared on a reasonable request to the corresponding author.

%%%%%%%%%%%%%%%%%%%% REFERENCES %%%%%%%%%%%%%%%%%%

% The best way to enter references is to use BibTeX:

\bibliographystyle{mnras}
\bibliography{VoidLensing_v3} % if your bibtex file is called example.bib

% Don't change these lines
\bsp	% typesetting comment

\label{lastpage}
\end{document}